\documentclass[letterpaper,twocolumn,10pt]{article}

\usepackage[utf8]{inputenc}
\usepackage[english]{babel}
\usepackage{xspace}
\usepackage[T1]{fontenc}

\usepackage[hyphens]{url}
\usepackage[normalem]{ulem}
\usepackage[dvipsnames, table]{xcolor}

\usepackage{authblk}

\usepackage{usenix2019_v3}

\usepackage{multirow, multicol, booktabs, tabulary, tabu, longtable, array, varwidth}
\usepackage{placeins, lipsum}
\usepackage[flushleft]{threeparttable}
\usepackage{tablefootnote}

\usepackage{graphicx}
\usepackage[labelfont=bf, skip=2pt]{caption}
\usepackage{subcaption}
\usepackage{tikz, epstopdf, stfloats, bbding, capt-of}
\usepackage{algorithm}
\usepackage{algpseudocode}
\usepackage{mwe}  % example image

\usepackage{enumerate}

\usepackage[inline]{enumitem}
\setlist{noitemsep,nolistsep,leftmargin=*}

\usepackage{amsmath, amsfonts, amssymb, nicefrac}
\usepackage{amsthm}
\usepackage{siunitx}

\usepackage{algorithm}
\usepackage{algpseudocode}

\usepackage[]{cleveref} % get fancy referencing
\crefname{appsec}{Appendix}{Appendices}
\crefname{definition}{Def.}{Defs.}
\crefformat{section}{\S#2#1#3}
\crefrangeformat{section}{\S#3#1#4--\S#5#2#6}
\crefformat{subsection}{\S#2#1#3}
\crefformat{subsubsection}{\S#2#1#3}
\crefrangeformat{subsection}{\S#3#1#4--\S#5#2#6}
\crefmultiformat{subsection}{\S#2#1#3}{ and~\S#2#1#3}{, \S#2#1#3}{ and~\S#2#1#3}
\crefformat{equation}{(#2#1#3)}
\crefrangeformat{equation}{(#3#1#4--#5#2#6)}
\crefmultiformat{equation}{(#2#1#3)}{ and~(#2#1#3)}{, (#2#1#3)}{ and~(#2#1#3)}
\crefformat{figure}{Fig.~#2#1#3}
\crefrangeformat{figure}{Figs. #3#1#4--#5#2#6}
\crefmultiformat{figure}{Figs.~#2#1#3}{ and~#2#1#3}{, #2#1#3}{ and~#2#1#3}
\crefformat{algorithm}{Alg.~#2#1#3}
\crefformat{table}{Table~#2#1#3}
\crefrangeformat{table}{Tables~#3#1#4--#5#2#6}
\crefmultiformat{table}{Tables~#2#1#3}{ and~#2#1#3}{, #2#1#3}{ and~#2#1#3}

\usepackage[compact]{titlesec}
\titlespacing*{\section}{0pt}{6pt plus 4pt minus 2pt}{2pt plus 2pt minus 2pt}
\titlespacing*{\subsection}{0pt}{4pt plus 2pt minus 1pt}{2pt plus 1pt minus 1pt}
\titlespacing*{\subsubsection}{0pt}{4pt plus 2pt minus 1pt}{2pt plus 1pt minus 1pt}

\usepackage{accents}

\newcommand{\xxx}{FIRM\xspace}

\newcommand*\circled[1]{\textcolor{red!70!black}{\tikz[baseline=(char.base)]{\node[shape=circle,draw,inner sep=1pt,thick] (char) {\textbf{#1}};}}}

\theoremstyle{definition}
\newtheorem{definition}{Definition}[section]

\newcommand{\uhat}{\underaccent{\check}}

\newcommand*{\affaddr}[1]{\textit{#1}} % No op here. Customize it for different styles.
\newcommand*{\affmark}[1][*]{\textsuperscript{#1}}

\begin{document}
%%%%%%%%%%%%%%%%%%%%%%%%%%%%%%%%%%%%%%%%%%%%%%%%%%%%%%%%%%%%%%%%%%%%%%%%%%

\title{\Large \bf \xxx: An Intelligent \underline{Fi}ne-Grained \underline{R}esource \underline{M}anagement Framework \\ for SLO-Oriented Microservices}

%\author{\# 258}
\author{
Haoran Qiu\affmark[1] \hspace{4pt} Subho S. Banerjee\affmark[1] \hspace{4pt} Saurabh Jha\affmark[1] \\ Zbigniew T. Kalbarczyk\affmark[2] \hspace{4pt} Ravishankar K. Iyer\affmark[1,2]
\vspace{6pt}\\
\affaddr{\affmark[1]Department of Computer Science} \hspace{4pt} \affaddr{\affmark[2]Department of Electrical and Computer Engineering}\\
\affaddr{University of Illinois at Urbana-Champaign}%\\
%\email{\{haoranq4,ssbaner2,sjha8,kalbarcz,rkiyer\}@illinois.edu}
}
\maketitle

%%%%%%%%%%%%%%%%%%%%%%%%%%%%%%%%%%%%%%%%%%%%%%%%%%%%%%%%%%%%%%%%%%%%%%%%%%
    \begin{abstract}
User-facing latency-sensitive web services include numerous distributed, intercommunicating microservices that promise to simplify software development and operation.
However, multiplexing of compute resources across microservices is still challenging in production because contention for shared resources can cause latency spikes that violate the service-level objectives (SLOs) of user requests.
This paper presents \textit{\xxx}, an intelligent fine-grained resource management framework for predictable sharing of resources across microservices to drive up overall utilization.
\xxx leverages online telemetry data and machine-learning methods to adaptively
\begin{enumerate*}[label=(\alph*)]
    \item detect/localize microservices that cause SLO violations,
    \item identify low-level resources in contention, and
    \item take actions to mitigate SLO violations via dynamic reprovisioning.
\end{enumerate*}
Experiments across four microservice benchmarks demonstrate that \xxx reduces SLO violations by up to 16$\times$ while reducing the overall requested CPU limit by up to 62\%.
Moreover, \xxx improves performance predictability by reducing tail latencies by up to 11$\times$.
\end{abstract}
    \section{Introduction}
\label{sec:intro}
User-facing latency-sensitive web services, like those at Netflix~\cite{microservices:netflix}, Google~\cite{microservices:google}, and Amazon~\cite{microservices:amazon}, are increasingly built as microservices that execute on shared/multi-tenant compute resources either as virtual machines (VMs) or as containers (with containers gaining significant popularity of late).
These microservices must handle diverse load characteristics while efficiently multiplexing shared resources in order to maintain service-level objectives (SLOs) like end-to-end latency.
SLO violations occur when one or more ``critical'' microservice instances (defined in \cref{sec:background}) experience load spikes (due to diurnal or unpredictable workload patterns) or shared-resource contention, both of which lead to longer than expected times to process requests, i.e., latency spikes~\cite{ding2019characterizing,dean2013tail,barroso2009datacenter,sriraman2018mu,sriraman2018mutune,gan2019seer,jindal2019performance,arapakis2014impact,latencycost:money,latencycost:ranking}.
Thus, it is critical to efficiently multiplex shared resources among microservices to reduce SLO violations.

% This problem is hard because...
% normal approaches use high level abstractions 
% we show significant improvement by leveraging low-level resources and using a multilevel ML approach

Traditional approaches (e.g., overprovisioning~\cite{reiss2012heterogeneity,ganguli2018cpu}, recurrent provisioning~\cite{jyothi2016morpheus,marr2016automated}, and autoscaling~\cite{qu2018auto,lorido2014review,yu2019microscaler,gias2019atom,prachitmutita2018auto,kalavri2018three,rzadca2020autopilot}) reduce SLO violations by allocating more CPUs and memory to microservice instances by using performance models, handcrafted heuristics (i.e., static policies), or machine-learning algorithms.

\begin{figure}[!t]
    \centering
    \includegraphics[width=\columnwidth]{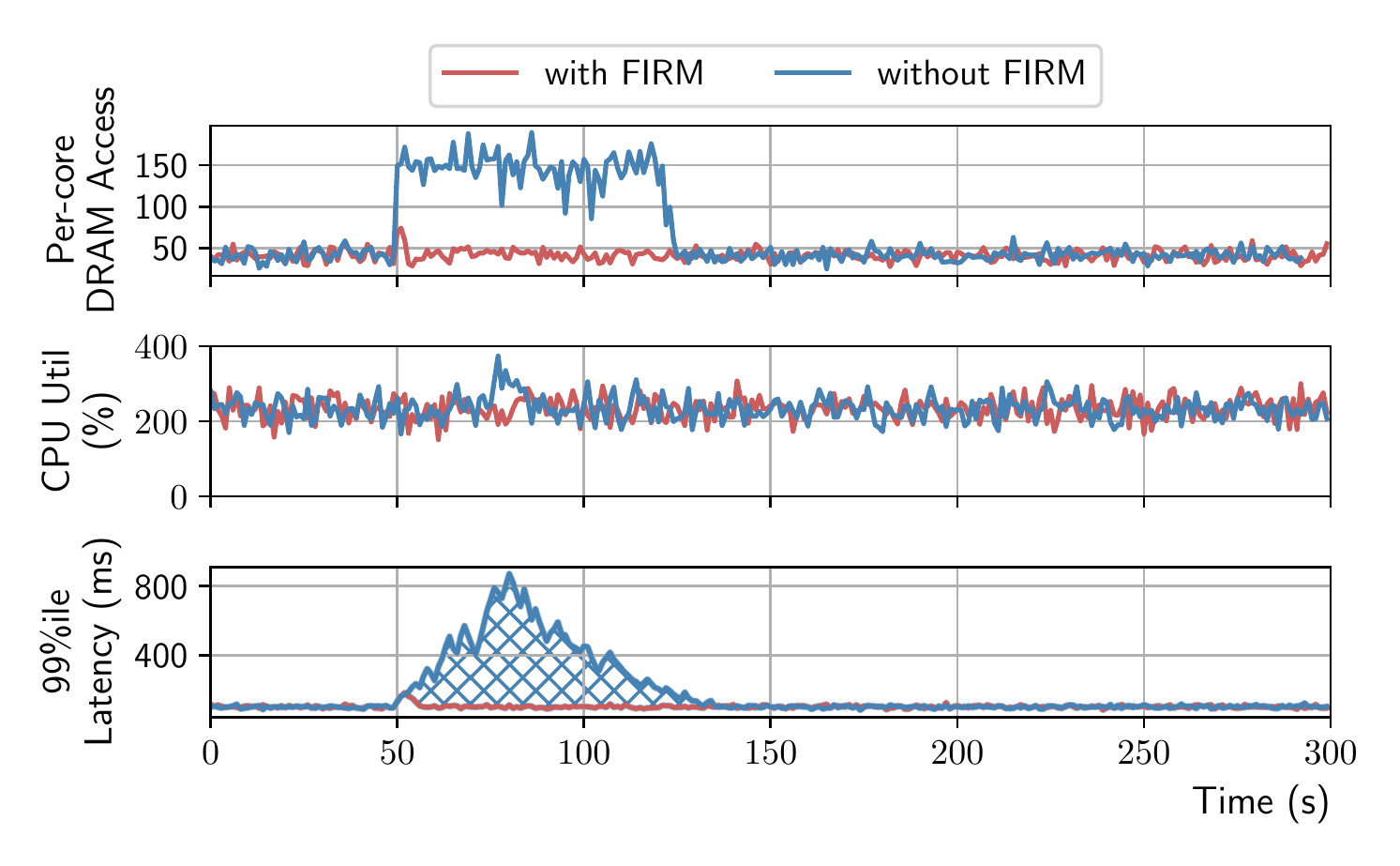}
    \vspace{-20pt}
    \caption{Latency spikes on microservices due to low-level resource contention.}
    \label{fig:eval-intro}
\end{figure}

Unfortunately, these approaches suffer from two main problems.
First, they fail to efficiently multiplex resources, such as caches, memory, I/O channels, and network links, at fine granularity, and thus may not reduce SLO violations.
For example, in \cref{fig:eval-intro}, the Kubernetes container-orchestration system~\cite{container:k8s} is unable to reduce the tail latency spikes arising from contention for a shared resource like memory bandwidth, as its autoscaling algorithms were built using heuristics that only monitor CPU utilization, which does not change much during the latency spike.
Second, significant human-effort and training are needed to build high-fidelity performance models (and related scheduling heuristics) of large-scale microservice deployments (e.g., queuing systems~\cite{gias2019atom,delimitrou2018amdahl}) that can capture low-level resource contention.
Further, frequent microservice updates and migrations can lead to recurring human-expert-driven engineering effort for model reconstruction.

% cloud under-utilization problem
%Over-provisioning~\cite{reiss2012heterogeneity} lead to low resource utilization in datacenters~\cite{reiss2012heterogeneity,barroso2009datacenter,barroso2011warehouse,delimitrou2013ibench,delimitrou2014quasar}.
%On the other hand, the resource utilization of modern datacenters is relatively low~\cite{reiss2012heterogeneity,barroso2009datacenter,barroso2011warehouse,delimitrou2013ibench,delimitrou2014quasar}.
%It is reported that the utilization of CPU and memory in Google's datacenters are only about 35\% and 55\% respectively~\cite{reiss2011google}, implying that a large amount of the installed resources cannot be fully utilized.

\textbf{\xxx Framework.}
This paper addresses the above problems by presenting \textit{\xxx}, a multilevel machine learning (ML) based resource management (RM) framework to manage shared resources among microservices at finer granularity to reduce resource contention and thus increase performance isolation and resource utilization.
As shown in \cref{fig:eval-intro}, \xxx performs better than a default Kubernetes autoscaler because \xxx adaptively scales up the microservice (by adding local cores) to increase the aggregate memory bandwidth allocation, thereby effectively maintaining the per-core allocation.
\xxx leverages online telemetry data (such as request-tracing data and hardware counters) to capture the system state, and ML models for resource contention estimation and mitigation.
Online telemetry data and ML models enable \xxx to adapt to workload changes and alleviate the need for brittle, handcrafted heuristics.
In particular, \xxx uses the following ML models:
\begin{itemize}
    \item \textit{Support vector machine (SVM) driven detection and localization of SLO violations to individual microservice instances}.
    \xxx first identifies the ``critical paths,'' and then uses per-critical-path and per-microservice-instance performance variability metrics (e.g., sojourn time~\cite{adan2002queueing}) to output a binary decision on whether or not a microservice instance is responsible for SLO violations.
    \item \textit{Reinforcement learning (RL) driven mitigation of SLO violations that reduces contention on shared resources}.
    \xxx then uses resource utilization, workload characteristics, and performance metrics to make dynamic reprovisioning decisions, which include 
    \begin{enumerate*}[label=(\alph*)]
        \item increasing or reducing the partition portion or limit for a resource type,
        \item scaling up/down, i.e., adding or reducing the amount of resources attached to a container, and
        \item scaling out/in, i.e., scaling the number of replicas for services.
    \end{enumerate*}
    By continuing to learn mitigation policies through reinforcement, \xxx can optimize for dynamic workload-specific characteristics.
\end{itemize}

\textbf{Online Training for \xxx.}
We developed a \textit{performance anomaly injection framework} that can artificially create resource scarcity situations in order to both train and assess the proposed framework. 
The injector is capable of injecting resource contention problems at a fine granularity (such as last-level cache and network devices) to trigger SLO violations.
To enable rapid (re)training of the proposed system as the underlying systems~\cite{mars2013whare} and workloads~\cite{sriraman2020accelerometer,sriraman2018mu,hazelwood2018applied,gmach2007workload} change in datacenter environments, \xxx uses \textit{transfer learning}.
That is, \xxx leverages transfer learning to train microservice-specific RL agents based on previous RL experience.

\textbf{Contributions.}
To the best of our knowledge, this is the first work to provide an SLO violation mitigation framework for microservices by using fine-grained resource management in an application-architecture-agnostic way with multilevel ML models.
Our main contributions are:
\begin{enumerate}
    % \item \emph{Characterization:} We characterize (in \cref{sec:background}) the impact of shared-resource contention for a set of real-world, latency-critical benchmark workloads~\cite{gan2019open, traintickets}.
    % We show that the impact of low-level resource contention destroys performance isolation and is highly workload dependent, thus precluding the possibility of using static policies for resource partitioning, and demanding management over fine-grained shared-resources instead of only scaling CPU or memory.
    % % Fine-grained resource contention problems cannot be solved by scaling CPU or memory.
    % We also show that different causes of SLO violation manifest in evolving critical paths, which lead to significant latency variations.
    
    \item \emph{SVM-based SLO Violation Localization:} 
    We present (in \cref{sec:critical-path,sec:critical-component}) an efficient way of localizing the microservice instances responsible for SLO violations by extracting critical paths and detecting anomaly instances in near-real time using telemetry data.
   
    \item \emph{RL-based SLO Violation Mitigation:} We present (in \cref{sec:mitigation}) an RL-based resource contention mitigation mechanism that (a) addresses the large state space problem and (b) is capable of tuning tailored RL agents for individual microservice instances by using transfer learning.
    % We use an actor-critic method DDPG~\cite{lillicrap2015continuous} with parameter noise to cater to large state space problem and help RL agent exploration.
    
    \item \emph{Online Training \& Performance Anomaly Injection:} We propose (in \cref{sec:fault-injector}) a comprehensive performance anomaly injection framework to artificially create resource contention situations, thereby generating the ground-truth data required for training the aforementioned ML models. 
    
    \item \emph{Implementation \& Evaluation:} We provide an open-source implementation of \xxx for the Kubernetes container-orchestration system~\cite{container:k8s}. We demonstrate and validate this implementation on four real-world microservice benchmarks~\cite{gan2019open,traintickets} (in \cref{sec:eval}).
\end{enumerate}

\textbf{Results.}
\xxx significantly outperforms state-of-the-art RM frameworks like Kubernetes autoscaling~\cite{container:k8s,k8sautoscaling} and additive increase multiplicative decrease (AIMD) based methods~\cite{studli2015modified,gevros2004distributed}.
\begin{itemize}
    \item It reduces overall SLO violations by up to 16$\times$ compared with Kubernetes autoscaling, and 9$\times$ compared with the AIMD-based method, while reducing the overall requested CPU by as much as 62\%.
    \item It outperforms the AIMD-based method by up to 9$\times$ and Kubernetes autoscaling by up to 30$\times$ in terms of the time to mitigate SLO violations.
    \item It improves overall performance predictability by reducing the average tail latencies up to 11$\times$.
    \item It successfully localizes SLO violation root-cause microservice instances with 93\% accuracy on average.
\end{itemize}

%\textbf{Why does \xxx work?}
\xxx mitigates SLO violations without overprovisioning because of two main features.
First, it models the dependency between low-level resources and application performance in an RL-based feedback loop to deal with uncertainty and noisy measurements.
Second, it takes a two-level approach in which the online critical path analysis and the SVM model filter only those microservices that need to be considered to mitigate SLO violations, thus making the framework application-architecture-agnostic as well as enabling the RL agent to be trained faster.

    \begin{figure*}[!t]
    \centering
    \includegraphics[width=\linewidth]{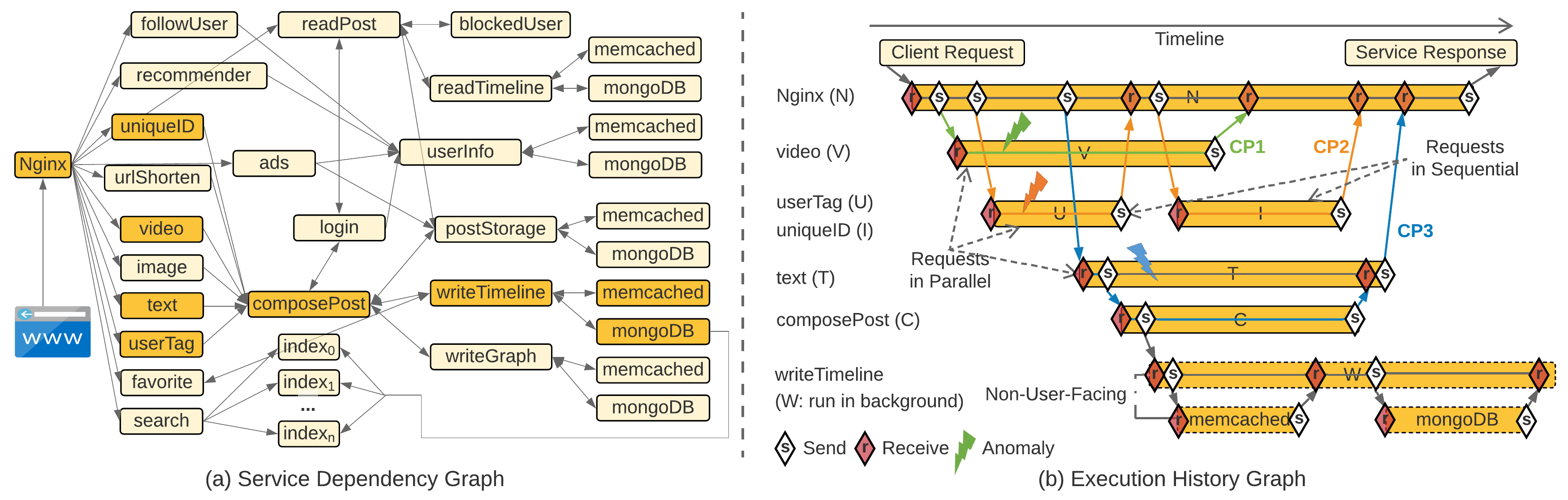}
    \vspace{-.4cm}
    \caption{Microservices overview: (a) Service dependency graph of \textit{Social Network} from the DeathStarBench~\cite{gan2019open} benchmark; (b) Execution history graph of a \texttt{post-compose} request in the same microservice.}
    \label{fig:microservice-overview}
\end{figure*}

\begin{figure*}[!t]
    \begin{subfigure}[b]{0.25\textwidth}
        \centering
        \includegraphics[width=\linewidth]{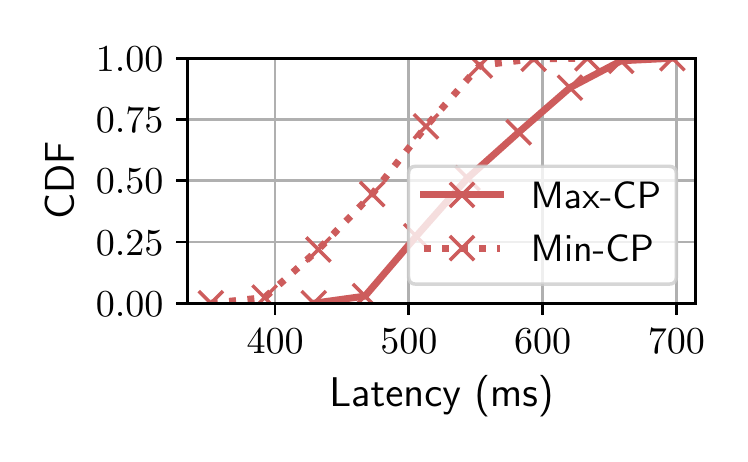}
        \vspace{-20pt}
        \caption{Social network service.}
    \end{subfigure}%
    \hfill%
    \begin{subfigure}[b]{0.25\textwidth}
        \centering
        \includegraphics[width=\linewidth]{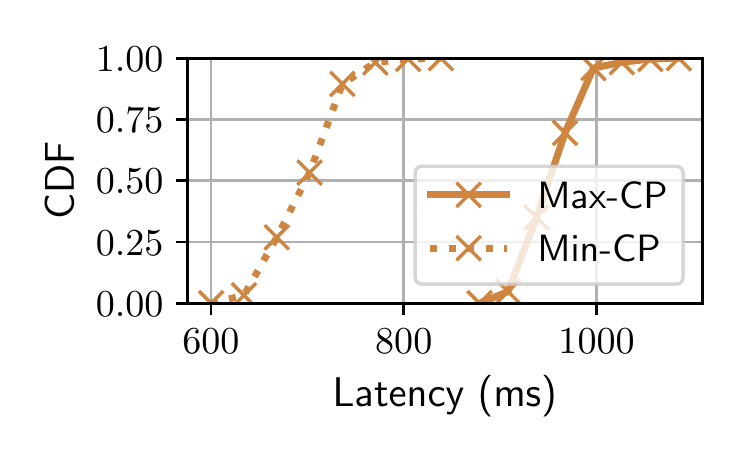}
        \vspace{-20pt}
        \caption{Media service.}
    \end{subfigure}%
    \hfill%
    \begin{subfigure}[b]{0.25\textwidth}
        \centering
        \includegraphics[width=\linewidth]{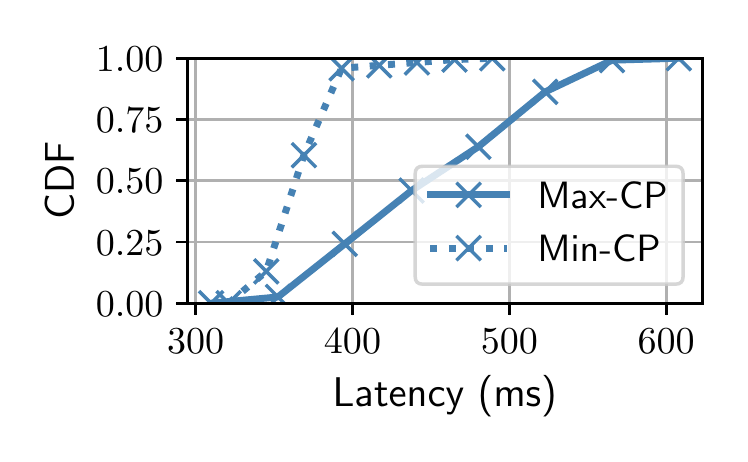}
        \vspace{-20pt}
        \caption{Hotel reservation service.}
    \end{subfigure}%
    \hfill%
    \begin{subfigure}[b]{0.25\textwidth}
        \centering
        \includegraphics[width=\linewidth]{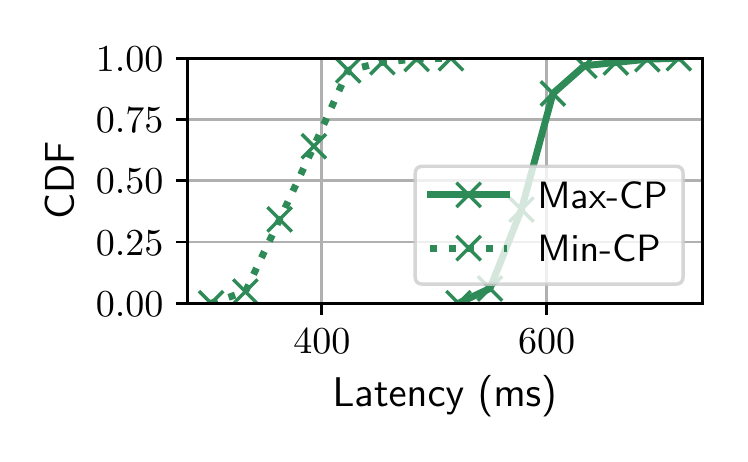}
        \vspace{-20pt}
        \caption{Train-ticket booking service.}
    \end{subfigure}
    \caption{Distributions of end-to-end latencies of different microservices in the DeathStarBench~\cite{gan2019open} and Train-Ticket~\cite{traintickets} benchmarks. Dashed and solid lines correspond to the minimum and maximum critical path latencies on serving a request.}
    \label{fig:cp-differences}
\end{figure*}

\section{Background \& Characterization}
\label{sec:background}

The advent of \textit{microservices} has led to the development and deployment of many web services that are composed of ``micro,'' loosely coupled, intercommunicating services, instead of large, monolithic designs.
This increased popularity of service-oriented architectures (SOA) of web services has been made possible by the rise of containerization~\cite{senthil2017practical,merkel2014docker,container:rkt,thalheim2018cntr} and container-orchestration frameworks~\cite{container:dockerswarm,container:k8s,schwarzkopf2013omega,verma2015large} that enable modular, low-overhead, low-cost, elastic, and high-efficiency development and production deployment of SOA microservices~\cite{microservices:netflix,microservices:linkedin,microservices:google,microservices:amazon,balalaie2015migrating,balalaie2016microservices,taibi2017processes,gan2018architectural,gan2019open}.
A deployment of such microservices can be visualized as a \emph{service dependency graph} or an \emph{execution history graph}.
The performance of a user request, i.e., its end-to-end latency, is determined by the \emph{critical path} of its execution history graph. 

\begin{definition}
    A \textit{service dependency graph} captures communication-based dependencies (the edges of the graph) between microservice instances (the vertices of the graph), such as remote procedure calls (RPCs).
    It tells how requests are flowing among microservices by following parent-child relationship chains.
    \cref{fig:microservice-overview}(a) shows the service dependency graph of the \textit{Social Network} microservice benchmark~\cite{gan2019open}. 
    Each user request traverses a subset of vertices in the graph. 
    For example, in \cref{fig:microservice-overview}(a), \texttt{post-compose} requests traverse only those microservices highlighted in darker yellow.
\end{definition}

\begin{definition}
    An \textit{execution history graph} is the space-time diagram of the distributed execution of a user request, where a vertex is one of \texttt{send\_req}, \texttt{recv\_req}, and \texttt{compute}, and edges represent the RPC invocations corresponding to \texttt{send\_req} and \texttt{recv\_req}.
    The graph is constructed using the global view of execution provided by distributed tracing of all involved microservices.
    For example, \cref{fig:microservice-overview}(b) shows the execution history graph for the user request in \cref{fig:microservice-overview}(a).
\end{definition}

\begin{definition}
    \label{def:cp}
    The \textit{critical path} (CP) to a microservice $m$ in the execution history graph of a request is the path of maximal duration that starts with the client request and ends with $m$~\cite{lockyer1969introduction,yang1988critical}.
    When we mention CP alone without the target microservice $m$, it means the critical path of the ``Service Response'' to the client (see \cref{fig:microservice-overview}(b)), i.e., end-to-end latency.
\end{definition}

To understand SLO violation characteristics and study the relationship between runtime performance and the underlying resource contention,
we have run extensive performance anomaly injection experiments on widely used microservice benchmarks (i.e. DeathStarBench~\cite{gan2019open} and Train-Ticket~\cite{traintickets}) and collected around 2 TB of raw tracing data (over $4.1 \times 10^7$ traces).
Our key insights are as follows.

\textbf{Insight 1: Dynamic Behavior of CPs.}
In microservices, the latency of the CP limits the overall latency of a user request in a microservice.
However, CPs do not remain static over the execution of requests in microservices, but rather change dynamically based on the performance of individual service instances because of underlying shared-resource contention and their sensitivity to this interference. Though other causes may also lead to CP evolution at real-time (e.g., distributed rate limiting~\cite{raghavan2007cloud}, and cacheability of requested data~\cite{ager2010revisiting}), it can still be used as an efficient manifestation of resource interference.

For example, in \cref{fig:microservice-overview}(b), we show the existence of three different CPs (i.e., CP1--CP3) depending on which microservice (i.e., $V$, $U$, $T$) encounters resource contention. 
We artificially create resource contention by using \textit{performance anomaly injections}.\footnote{\textit{Performance anomaly injections} (\cref{sec:fault-injector}) are used to trigger SLO violations by generating fine-grained resource contention with configurable resource types, intensity, duration, timing, and patterns, which helps with both our characterization (\cref{sec:background}) and ML model training (\cref{sec:mitigation}).}
\cref{tab:critical-path-changes} lists the changes observed in the latencies of individual microservices, as well as end-to-end latency.
We observe as much as 1.2--2$\times$ variation in end-to-end latency across the three CPs.
Such dynamic behavior exists across all our benchmark microservices.
\cref{fig:cp-differences} illustrates the latency distributions of CPs with minimum and maximum latency in each microservice benchmark, where we observe as much as 1.6$\times$ difference in median latency and 2.5$\times$ difference in 99th percentile tail latency across these CPs.

\begin{table}[!t]
    \centering
    \caption{CP changes in \cref{fig:microservice-overview}(b) under performance anomaly injection. Each case is represented by a <$service, CP$> pair. $N$, $V$, $U$, $I$, $T$, and $C$ are microservices from \cref{fig:microservice-overview}.}
    \label{tab:critical-path-changes}
    \resizebox{\columnwidth}{!}{%
        \begin{tabular}{lccccccr}
        \toprule
        \multirow{2}{*}{\textbf{Case}} & \multicolumn{6}{c}{\centering \textbf{Average Individual Latency ($ms$)}} & \multirow{2}{*}{\textbf{Total ($ms$)}} \\
        \cmidrule(l){2-7}
        & $N$ & $V$ & $U$ & $I$ & $T$ & $C$\\
        \midrule
        <$V, CP1$> & 13 & 603 & 166 & 33 & 71 & 68 & 614 $\pm$ 106\\
        <$U, CP2$> & 14 & 237 & 537 & 39 & 62 & 89 & 580 $\pm$ 113\\
        <$T, CP3$> & 13 & 243 & 180 & 35 & 414 & 80 & 507 $\pm$ 75\\
        \bottomrule
        \end{tabular}%
    }
\end{table}

\begin{figure}[!t]
    \begin{subfigure}[b]{0.235\textwidth}
        \centering
        \includegraphics[width=\linewidth]{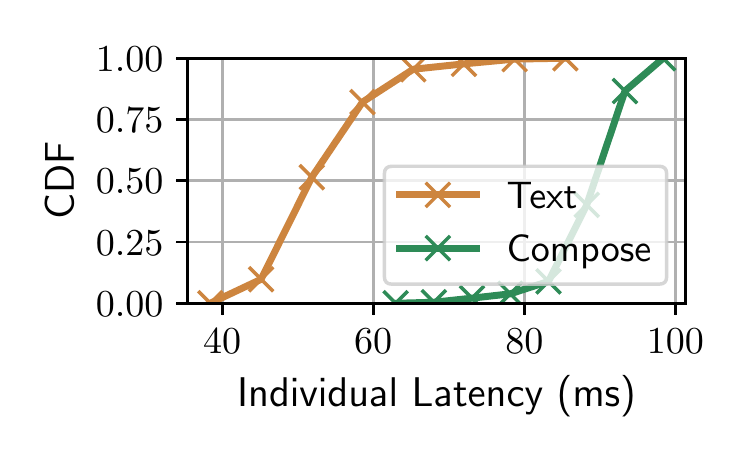}
        \vspace{-17pt}
    \end{subfigure}%
    \hfill%
    \begin{subfigure}[b]{0.235\textwidth}
        \centering
        \includegraphics[width=\linewidth]{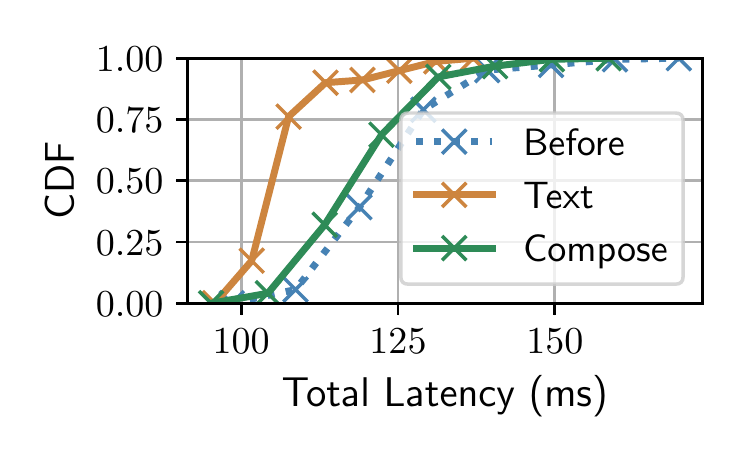}
        \vspace{-17pt}
    \end{subfigure}%
    \caption{Improvement of end-to-end latency by scaling ``highest-variance'' and ``highest-median'' microservices.}
    \label{fig:h0ha}
\end{figure}

Recent approaches (e.g.,~\cite{ilyushkin2017experimental,ahn2014auto}) have explored static identification of CPs based on historic data (profiling) and have built heuristics (e.g., application placement, level of parallelism) to enable autoscaling to minimize CP latency.
However, our experiment shows that this by itself is not sufficient.
The requirement is to \textit{adaptively capture changes in the CPs}, in addition to changing resource allocations to microservice instances on the identified CPs to mitigate tail latency spikes.

\textbf{Insight 2: Microservices with Larger Latency Are Not Necessarily Root Causes of SLO Violations.}
It is important to find the microservices responsible for SLO violations to mitigate them.
While it is clear that such microservices will always lie on the CP, it is less clear which individual service on the CP is the culprit.
A common heuristic is to pick the one with the highest latency.
However, we find that that rarely leads to the optimal solution.
Consider \cref{fig:h0ha}. The left side shows the CDF of the latencies of two services (i.e., \texttt{composePost} and \texttt{text}) on the CP of the \texttt{post-compose} request in the Social Network benchmark.
The \texttt{composePost} service has a higher median/mean latency while the \texttt{text} service has a higher variance.
Now, although the \texttt{composePost} service contributes a larger portion of the total latency, it does not benefit from scaling (i.e., getting more resources), as it does not have resource contention.
That phenomenon is shown on the right side of \cref{fig:h0ha}, which shows the end-to-end latency for the original configuration (labeled ``Before'') and after the two microservices were scaled from a single to two containers each (labeled ``Text'' and ``Compose''). Hence, scaling microservices with higher variances provides better performance gain.

\begin{figure}[!t]
    \centering
    \includegraphics[width=\linewidth]{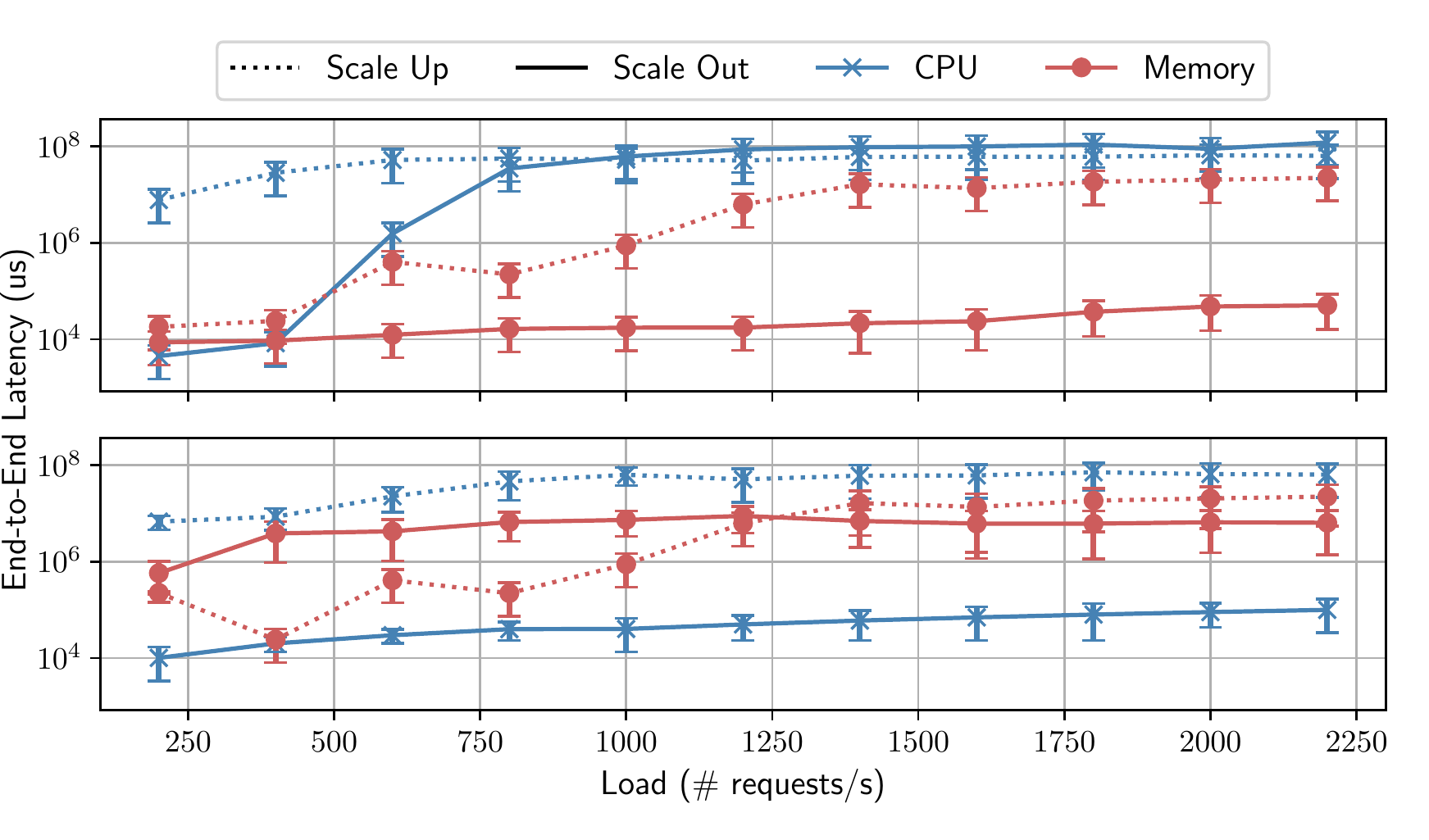}
    \caption{Dynamic behavior of mitigation strategies: \textit{Social Network} (top); \textit{Train-Ticket Booking} (bottom). Error bars show 95\% confidence intervals on median latencies.}
    \label{fig:variability}
\end{figure}

\textbf{Insight 3: Mitigation Policies Vary with User Load and Resource in Contention.}
The only way to mitigate the effects of dynamically changing CPs, which in turn cause dynamically changing latencies and tail behaviors, is to efficiently identify microservice instances on the CP that are resource-starved or contending for resources and then provide them with more of the resources.
Two common ways of doing so are 
\begin{enumerate*}[label=(\alph*)]
    \item to \textit{scale out} by spinning up a new instance of the container on another node of the compute cluster, or 
    \item to \textit{scale up} by providing more resources to the container via either explicitly partitioning resources (e.g., in the case of memory bandwidth or last-level cache) or granting more resources to an already deployed container of the microservice (e.g., in the case of CPU cores). 
\end{enumerate*}

As described before, recent approaches~\cite{qu2018auto,lorido2014review,yu2019microscaler,gias2019atom,kalavri2018three,studli2015modified,gevros2004distributed,dejun2011resource,sharma2011cost}) address the problem by building static policies (e.g., AIMD for controlling resource limits~\cite{studli2015modified,gevros2004distributed}, and rule/heuristics-based scaling relying on profiling of historic data about a workload~\cite{dejun2011resource,sharma2011cost}), or modeling performance~\cite{gias2019atom,kalavri2018three}.
% ML - prachitmutita2018auto,yang2019miras,rzadca2020autopilot
However, we found in our experiments with the four microservice benchmarks that such static policies are not well-suited for dealing with latency-critical workloads because the optimal policy must incorporate dynamic contextual information.
That is, information about the type of user requests, and load (in requests per second), as well as the critical resource bottlenecks (i.e, the resource being contended for), must be jointly analyzed to make optimal decisions.
For example, in \cref{fig:variability} (top), we observe that the trade-off between scale-up and scale-out changes based not only on the user load but also on the resource type. 
At 500 req/s, scale-up has a better payoff (i.e, lower latency) than scale-out for both memory- and CPU-bound workloads. 
However, at 1500 req/s, scale-out dominates for CPU, and scale-up dominates for memory. 
This behavior is also application-dependent because the trade-off curve inflection points change across applications, as illustrated in \cref{fig:variability} (bottom).

% To motivate the design goals and principles of the \xxx framework in microservice orchestration and management, we demonstrate the potential pitfalls of the state of the art techniques in measurement-driven study of the Kubernetes container-orchestration system running a set of benchmark workloads~\cite{gan2019open, traintickets}.
    \section{The \xxx Framework}
\label{sec:overview} 

\begin{figure}[!t]
  \includegraphics[width=\linewidth]{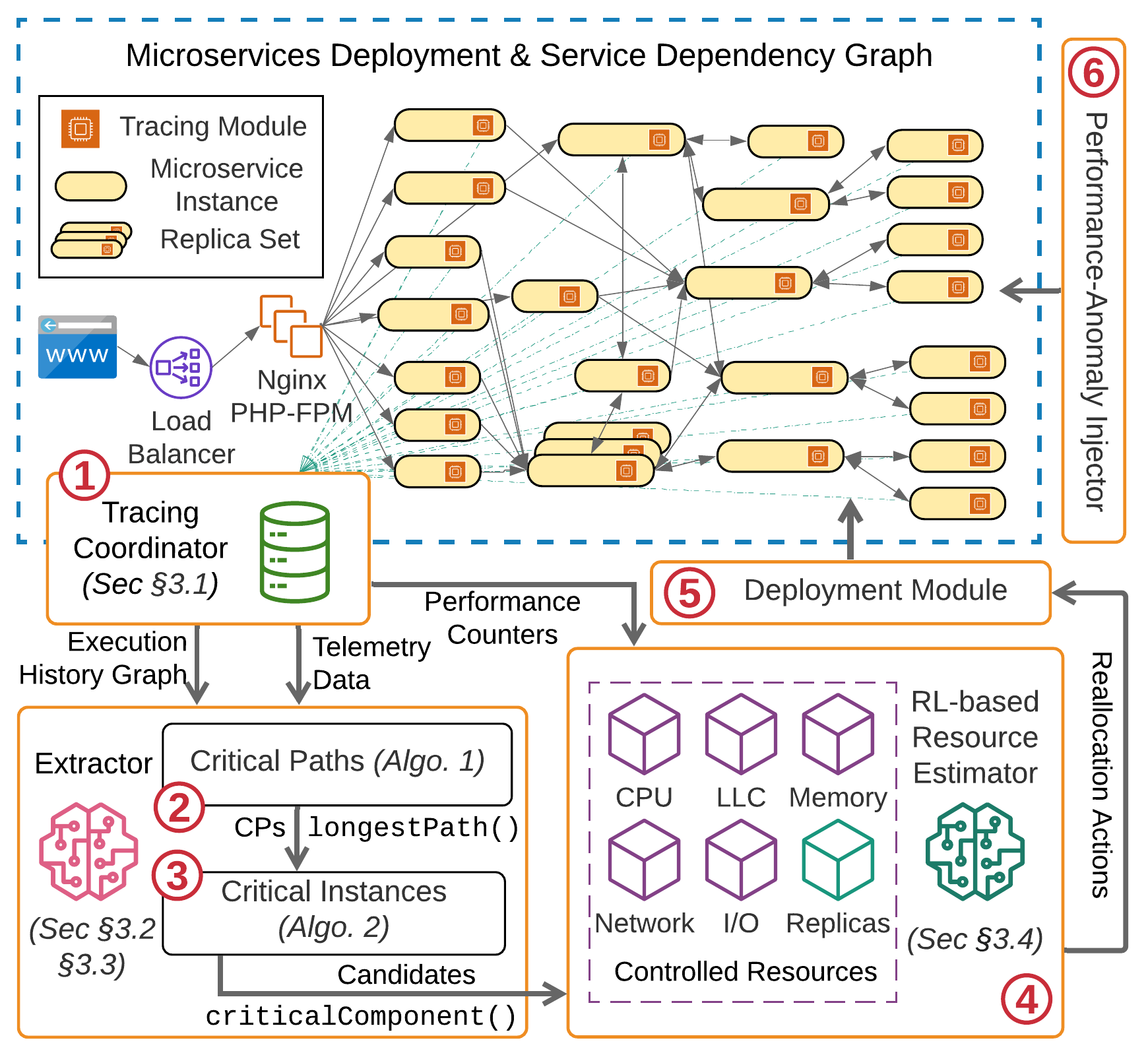}
  \caption{\xxx architecture overview.}
  \label{fig:overview}
\end{figure}

In this section, we describe the overall architecture of the \xxx framework and its implementation.
\begin{enumerate}
    \item Based on the insight that resource contention manifests as dynamically evolving CPs, \xxx first detects CP changes and extracts critical microservice instances from them.
    It does so using the \textit{Tracing Coordinator}, which is marked as \circled{1} in \cref{fig:overview}.\footnote{Unless otherwise specified, \circled{*} refers to annotations in \cref{fig:overview}.}
    The tracing coordinator collects tracing and telemetry data from every microservice instance and stores them in a centralized graph database for processing.
    It is described in \cref{sec:tracing}.
    \item The \textit{Extractor} detects SLO violations and queries the Tracing Coordinator with collected real-time data
    \begin{enumerate*}[label=(\alph*)]
        \item to extract CPs (marked as \circled{2} and described in \cref{sec:critical-path}) and
        \item to localize critical microservice instances that are likely causes of SLO violations (marked as \circled{3} and described in \cref{sec:critical-component}).
    \end{enumerate*}
    \item Using the telemetry data collected in \circled{1} and the critical instances identified in \circled{3}, \xxx makes mitigation decisions to scale and reprovision resources for the critical instances (marked as \circled{4}).
    The policy used to make such decisions is automatically generated using RL.
    The RL agent jointly analyzes contextual information about resource utilization (i.e., low-level performance counter data collected from the CPU, LLC, memory, I/O, and network), performance metrics (i.e, per-microservice and end-to-end latency distributions), and workload characteristics (i.e., request arrival rate and composition) and makes mitigation decisions.
    The RL model and setup are described in \cref{sec:mitigation}.
    \item Finally, actions are validated and executed on the underlying Kubernetes cluster through the deployment module (marked as \circled{5} and described in \cref{sec:action-execution}).
    %If the virtual machine or physical machine is out of resource, then the action will be replaced by a scale-out operation.
    %This module is described in \cref{sec:action-execution}.
    \item In order to train the ML models in the Extractor as well as the RL agent (i.e., to span the exploration-exploitation trade-off space), \xxx includes a performance anomaly injection framework that triggers SLO violations by generating resource contention with configurable intensity and timing.
    This is marked as \circled{6} and described in \cref{sec:fault-injector}.
\end{enumerate}

\subsection{Tracing Coordinator}
\label{sec:tracing}

Distributed tracing is a method used to profile and monitor microservice-based applications to pinpoint causes of poor performance~\cite{tracing-instana, tracing-jaeger, tracing-lightstep, tracing-skywalking, tracing-zipkin}.
A \textit{trace} captures the work done by each service along request execution paths, i.e., it follows the execution ``route'' of a request across microservice instances and records time, local profiling information, and RPC calls (e.g., source and destination services).
The execution paths are combined to form the \textit{execution history graph} (see \cref{sec:background}).
The time spent by a single request in a microservice instance is called its \textit{span}.
The span is calculated based on the time when a request arrives at a microservice and when its response is sent back to the caller.
Each span is the most basic single unit of work done by a microservice.

The \xxx tracing module's design is heavily inspired by Dapper~\cite{sigelman2010dapper} and its open-source implementations, e.g., Jaeger~\cite{tracing-jaeger} and Zipkin~\cite{tracing-zipkin}.
Each microservice instance is coupled with an OpenTracing-compliant~\cite{opentracing} \textit{tracing agent} that measures spans.
As a result, any new OpenTracing-compliant microservice can be integrated naturally into the \xxx tracing architecture.
The Tracing Coordinator, i.e., \circled{1}, is a stateless, replicable data-processing component that collects the spans of different requests from each tracing agent, combines them, and stores them in a graph database~\cite{neo4j} as the execution history graph.
The graph database allows us to easily store complex caller-callee relationships among microservices depending on request types, as well as to efficiently query the graph for critical path/component extraction (see \cref{sec:critical-path,sec:critical-component}).
Distributed clock drift and time shifting are handled using the Jaeger framework.
In addition, the Tracing Coordinator collects telemetry data from the systems running the microservices.
The data collected in our experiments is listed in \cref{tab:telemetry_data}.
The distributed tracing and telemetry collection overhead is indiscernible, i.e., we observed a $<$0.4\% loss in throughput and a $<$0.15\% loss in latency.
\xxx had a maximum CPU overhead of 4.6\% for all loads running in our experiments on the four benchmarks~\cite{gan2019open,traintickets}.
With \xxx, the network in/out traffic without sampling traces increased by 3.4\%/10.9\% (in bytes); the increase could be less in production environments with larger message sizes~\cite{liu2019jcallgraph}.

\begin{table}[!t]
    \centering
    \caption{Collected telemetry data and sources.}
    \label{tab:telemetry_data}
    \resizebox{\columnwidth}{!}{%
    \begin{tabular}{>{\ttfamily}p{\columnwidth}}
        \toprule
        \rowcolor{gray!15}
        \multicolumn{1}{c}{\textbf{cAdvisor}~\cite{cadvisor} \& \textbf{Prometheus}~\cite{prometheus}}\\
        cpu\_usage\_seconds\_total, memory\_usage\_bytes,\\
        fs\_write/read\_seconds, fs\_usage\_bytes, network\_transmit/receive\_bytes\_total, processes\\
        \rowcolor{gray!15}
        \multicolumn{1}{c}{\textbf{Linux perf subsystem}~\cite{perf}}\\
        offcore\_response.*.llc\_hit/miss.local\_DRAM,\\
        offcore\_response.*.llc\_hit/miss.remote\_DRAM\\
        \bottomrule
    \end{tabular}%
    }
\end{table}

\subsection{Critical Path Extractor}
\label{sec:critical-path}

The first goal of the \xxx framework is to quickly and accurately identify the CP based on the tracing and telemetry data described in the previous section.
Recall from \cref{def:cp} in \cref{sec:background} that a CP is the longest path in the request's execution history graph.
Hence, changes in the end-to-end latency of an application are often determined by the slowest execution of one or more microservices on its CP.

We identify the CP in an execution history graph by using \cref{algo:critical-path}, which is a weighted longest path algorithm proposed to retrieve CPs in the microservices context.
The algorithm needs to take into account the major communication and computation patterns in microservice architectures:
\begin{enumerate*}[label=(\alph*)]
    \item \textit{parallel},
    \item \textit{sequential}, and
    \item \textit{background}
\end{enumerate*}
workflows.
\begin{itemize}
    \item \textit{Parallel workflows} are the most common way of processing requests in microservices.
    They are characterized by child spans of the same parent span that overlap with each other in the execution history graph, e.g., $U$, $V$, and $T$ in \cref{fig:microservice-overview}(b).
    Formally, for two child spans $i$ with start time $st_{i}$ and end time $et_{i}$, and $j$ with $st_{j}, et_{j}$ of the same parent span $p$, they are called \textit{parallel} if $(st_{j}<st_{i}<et_{j}) ~\lor~ (st_{i}<st_{j}<et_{i})$.

    \item \textit{Sequential workflows} are characterized by one or more child spans of a parent span that are processed in a serialized manner, e.g., $U$ and $I$ in \cref{fig:microservice-overview}(b).
    For two of $p$'s child-spans $i$ and $j$ to be in a sequential workflow, the time $t_{i\rightarrow p} \leq t_{p\rightarrow j}$, i.e., $i$ completes and sends its result to $p$ before $j$ does.
    Such sequential relationships are usually indicative of a \textit{happens-before} relationship.
    However, it is impossible to ascertain the relationships merely by observing traces from the system.
    If, across a sufficient number of request executions, there is a violation of that inequality, then the services are not sequential.

    \item \textit{Background workflows} are those that do not return values to their parent spans, e.g., $W$ in \cref{fig:microservice-overview}(b).
    Background workflows are not part of CPs since no other span depends on their execution, but they may be considered responsible for SLO violations when \xxx's Extractor is localizing critical components (see \cref{sec:critical-component}). That is because background workflows may also contribute to the contention of underlying shared resource.
\end{itemize}

\begin{algorithm}[!t]
\caption{Critical Path Extraction}
\label{algo:critical-path}
\begin{algorithmic}[1]
\Require Microservice execution history graph $G$
\Statex Attributes: $childNodes$, $lastReturnedChild$
\Procedure{LongestPath}{$G$, $currentNode$}
    \State $path \gets \varnothing$
    \State $path$.add($currentNode$)
    \If {$currentNode.childNodes$ == None}
        \State Return $path$
    \EndIf
    \State $lrc \gets currentNode.lastReturnedChild$
    \State $path$.extend(\textsc{LongestPath}($G$, $lrc$))
    \For {each $cn$ in $currentNode.childNodes$}
        \If {$cn$.happensBefore($lrc$)}
            \State $path$.extend(\textsc{LongestPath}($G$, $cn$))
        \EndIf
	\EndFor
	\State Return $path$
\EndProcedure
\end{algorithmic}
\end{algorithm}

\begin{algorithm}[!t]
\caption{Critical Component Extraction}
\label{algo:critical-component}
\begin{algorithmic}[1]
\Require Critical Path $CP$, Request Latencies $T$
\Procedure{CriticalComponent}{$G$, $T$}
    \State $candidates \gets \varnothing$
    \State $T_{CP} \gets T.getTotalLatency()$ \Comment{Vector of CP latencies}
    \For {$i \in CP$}
        \State $T_i \gets T.getLatency(i)$
        \State $T_{99} \gets T_i.percentile(99)$
        \State $T_{50} \gets T_i.percentile(50)$
        %\Comment{Pearson Correlation Coefficient}
        \State $RI \gets PCC(T_i, T_{CP})$ \Comment{Relative Importance}
        \State $CI \gets T_{99}/T_{50}$ \Comment{Congestion Intensity}
        \If {$SVM.classify(RI, CI)$ == $True$}
            \State $candidates.append(i)$
        \EndIf
	\EndFor
	\State Return $candidates$
\EndProcedure
\end{algorithmic}
\end{algorithm}

\subsection{Critical Component Extractor}
\label{sec:critical-component}

% \vspace{-4pt}
In each extracted CP, \xxx then uses an adaptive, data-driven approach to determine critical components (i.e., microservice instances).
The overall procedure is shown in \cref{algo:critical-component}.
The extraction algorithm first calculates per-CP and per-instance ``features,'' which represent the performance variability and level of request congestion.
Variability represents the single largest opportunity to reduce tail latency.
The two features are then fed into an incremental SVM classifier to get binary decisions, i.e., on whether that instance should have its resources re-provisioned or not.
The approach is a dynamic selection policy that is in contrast to static policies, as it can classify critical and noncritical components adapting to dynamically changing workload and variation patterns.
%We have run extensive (over $4.1 \times 10^7$ traces) performance anomaly injection experiments and collected > 2TB of microservice tracing data. 
%Our key observation is that high latency microservices are not necessarily the primary contributors to SLO violations.
%Instead, nearly all SLO violation cases are explained by microservices that have hight performance variability.
%\RED{Somehow quantify this observation: p-value? AB-test? Maybe there is a better metric.}

In order to extract those microservice instances that are potential candidates for SLO violations, we argue that it is critical to know both the variability of the end-to-end latency (i.e., per-CP variability) and the variability caused by congestion in the service queues of each individual microservice instances (i.e., per-instance variability).

\textbf{Per-CP Variability: Relative Importance.}
Relative importance~\cite{lindeman1980introduction,tonidandel2011relative,white2003algorithms} is a metric that quantifies the strength of the relationship between two variables.
For each critical path $CP$, its end-to-end latency is given by $T_{CP} = \sum_{i \in CP}{T_i}$, where $T_i$ is the latency of microservice $i$.
Our goal is to determine the contribution that the variance of each variable $T_i$ makes toward explaining the total variance of $T_{CP}$.
To do so, we use the Pearson correlation coefficient~\cite{benesty2009pearson} (also called zero-order correlation), i.e., $PCC(T_i, T_{CP})$, as the measurement, and hence the resulting statistic is known as the variance explained~\cite{eisinga2013reliability}.
The sum of $PCC(T_i, T_{CP})$ over all microservice instances along the CP is 1, and the relative importance values of microservices can be ordered by $PCC(T_i, T_{CP})$. The larger the value is, the more variability it contributes to the end-to-end CP variability.

\textbf{Per-Instance Variability: Congestion Intensity.}
For each microservice instance in a CP, congestion intensity is defined as the ratio of the 99th percentile latency to the median latency.
Here, we chose the 99th percentile instead of the 70th or 80th percentile to target the tail latency behavior.
The chosen ratio explains per-instance variability by capturing the congestion level of the request queue so that it can be used to determine whether it is necessary to scale.
For example, a higher ratio means that the microservice could handle only a subset of the requests, but the requests at the tail are suffering from congestion issues in the queue.
On the other hand, microservices with lower ratios handle most requests normally, so scaling does not help with performance gain.
Consequently, microservice instances with higher ratios have a greater opportunity to achieve performance gains in terms of tail latency by taking scale-out or reprovisioning actions.

\textbf{Implementation.}
The logic of critical path extraction is incorporated into the construction of spans, i.e., as the algorithm proceeds (\cref{algo:critical-path}), the order of tracing construction is also from the root node to child nodes recursively along paths in the execution history graph.
Sequential, parallel, and background workflows are inferred from the parent-child relationships of spans.
Then, for each CP, we calculate feature statistics and feed them into an incremental SVM classifier~\cite{diehl2003svm,laskov2006incremental} implemented using stochastic gradient descent optimization and RBF kernel approximation by \texttt{scikit-learn} libraries~\cite{scikit-learn}.
Triggered by detected SLO violations, both critical path extraction and critical component extraction are stateless and multithreaded; thus, the workload scales with the size of the microservice application and the cluster.
%warm-start active set algorithm~\cite{shilton2005incremental}.
%CP extraction (\textcolor{red!70!black}{\textbf{\circled{2}}}) and critical microservice instance extraction
They together constitute \xxx's extractor (i.e., \textcolor{red!70!black}{\textbf{\circled{2}}} and \textcolor{red!70!black}{\textbf{\circled{3}}}).
Experiments (\cref{eval:rca}) show that it reports SLO violation candidates with feasible accuracy and achieves completeness with \cref{sec:mitigation} by choosing a threshold with a reasonable false-positive rate.
% decision tree methods are prone to overfitting

\subsection{SLO Violation Mitigation Using RL}
\label{sec:mitigation}
%\vspace{-4pt}
Given the list of critical service instances, \xxx's Resource Estimator, i.e., \circled{4}, is designed to analyze resource contention and provide reprovisioning actions for the cluster manager to take.
\xxx estimates and controls a fine-grained set of resources, including CPU time, memory bandwidth, LLC capacity, disk I/O bandwidth, and network bandwidth.
It makes decisions on scaling each type of resource or the number of containers by using measurements of tracing and telemetry data (see \cref{tab:telemetry_data}) collected from the Tracing Coordinator.
When jointly analyzed, such data provides information about
\begin{enumerate*}[label=(\alph*)]
    \item shared-resource interference,
    \item workload rate variation, and
    \item request type composition.
\end{enumerate*}

\begin{figure}[!t]
    \centering
    \includegraphics[width=0.9\linewidth]{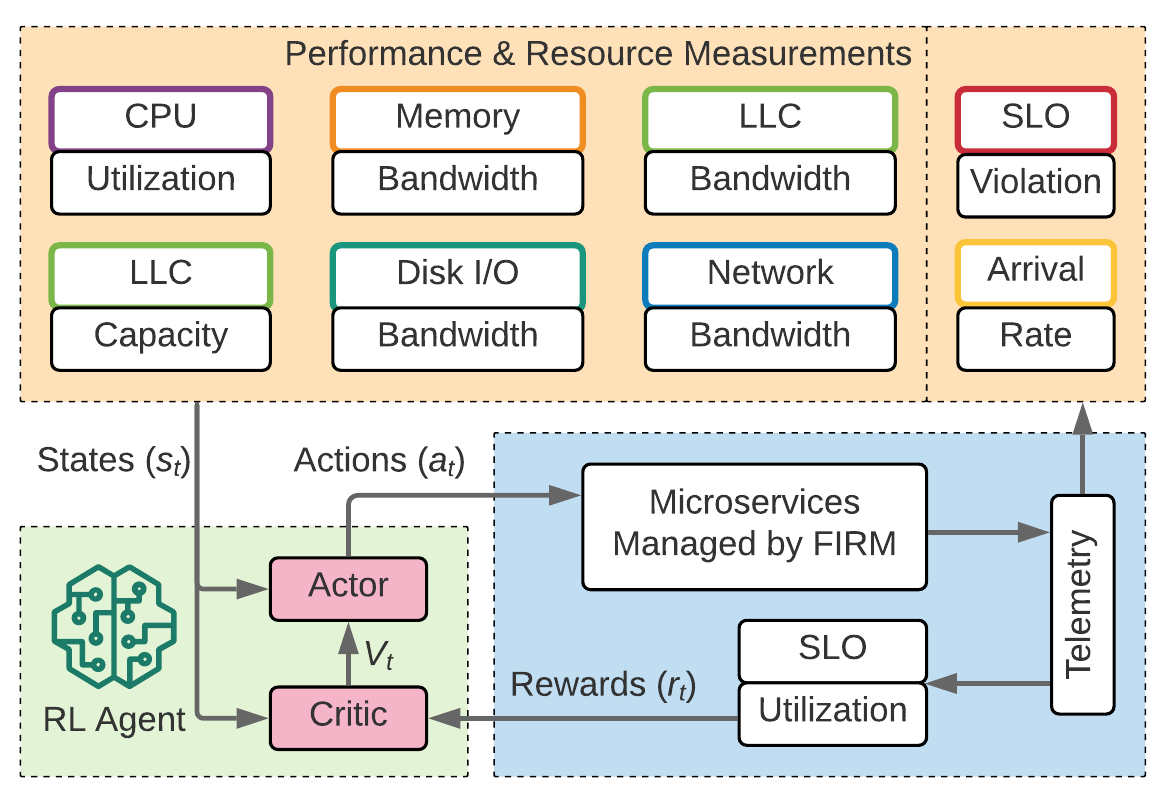}
    \caption{Model-free actor-critic RL framework for estimating resources in a microservice instance.}
    \label{fig:rl-agent}
\end{figure}

%The \xxx Resource Estimator works as follows.
%\begin{enumerate}
%    \item \xxx detects workload changes for each extracted critical service instance (i.e., an increase in the service instance request rate).
%    \item \xxx subjects anomalies that are not due to network delay to container resource contention identification (i.e., the RL method described below).
%\end{enumerate}
%Although containerization provides some degree of performance isolation to applications, shared resource contention at lower levels (particularly microarchitectural resources) can result in breaking isolation properties.
%This is because these low-level resources are shared by co-located containers, e.g., last-level cache, memory bandwidth, disk I/O and network devices.

% drawback of existing approaches
%However, existing work on resource estimation and allocation does not apply for microservices well.
%Performance modeling for microservices requires an elaborately crafted model depicting behavior of each microservice instance, but it is hard to get sophisticated prior knowledge of the microservices given the complexity and the scale.
%On the other hand, a data-driven modeling requires huge amount of training data from the running real microservices, which is resource and time-consuming.
%The resulting rule-based or heuristic-based approaches highly depend on complete understanding of the cloud microservices, and do not tackle with dynamic environments.

\xxx leverages reinforcement learning (RL) to optimize resource management policies for long-term reward in dynamic microservice environments.
% The feedback loop for training and inference is shown in \cref{fig:rl-agent}.
We next give a brief RL primer before presenting \xxx's RL model.

\textbf{RL Primer.}
An RL agent solves a \textit{sequential decision-making problem} (modeled as a Markov decision process) by interacting with an environment.
At each discrete time step $t$, the agent observes a \textit{state of the environment} $s_t \in S$, and performs an \textit{action} $a_t \in A$ based on its \textit{policy} $\pi_\theta(s)$ (parameterized by $\theta$), which maps \textit{state space} $S$ to \textit{action space} $A$.
At the following time step $t+1$, the agent observes an \textit{immediate reward} $r_t \in R$ given by a reward function $r(s_t, a_t)$; the immediate reward represents the loss/gain in transitioning from $s_t$ to $s_{t+1}$ because of action $a_t$.
The tuple ($s_t$,$a_t$,$r_t$,$s_{t+1}$) is called one \textit{transition}.
%The return from a state is defined as  $G_t = \sum_{k=0}^{\infty}\gamma^kr_{t+k}$
The agent's goal is to optimize the policy $\pi_\theta$ so as to maximize the
expected \textit{cumulative discounted reward} (also called the value function) from the start distribution $J = \mathbb{E}[G_1]$, where the return from a state $G_t$ is defined to be $\sum_{k=0}^{T}\gamma^kr_{t+k}$.
The discount factor $\gamma \in (0, 1]$ penalizes the predicted future rewards.

Two main categories of approaches are proposed for policy learning: value-based methods and policy based methods~\cite{arulkumaran2017brief}. In value-based methods, the agent learns an estimate of the optimal value function and approaches the optimal policy by maximizing it. In policy-based methods, the agent directly tries to approximate the optimal policy.

\begin{figure}[!t]
    \centering
    \includegraphics[width=0.9\linewidth]{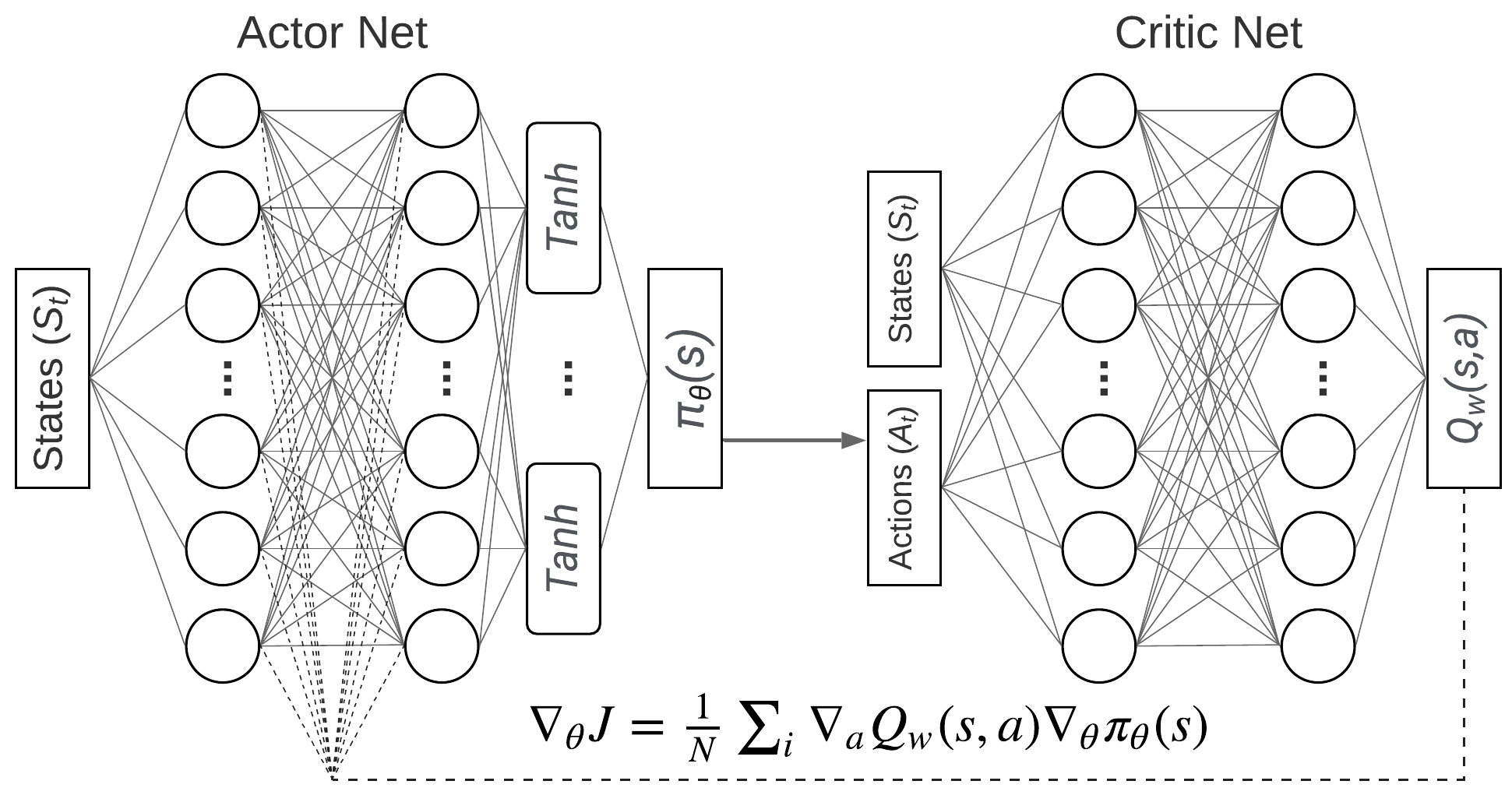}
    \caption{Architecture of actor-critic nets.}
    \label{fig:actor-critic-arch}
\end{figure}

\textbf{Why RL?}
Existing performance-modeling-based~\cite{yu2019microscaler,gias2019atom,kalavri2018three,studli2015modified,gevros2004distributed,dejun2011resource,sharma2011cost} or heuristic-based approaches~\cite{aws:autoscaling,azure:autoscaling,gcloud:autoscaling,qu2018auto,lorido2014review} suffer from model reconstruction and retraining problems because they do not address dynamic system status.
Moreover, they require expert knowledge, and it takes significant effort to devise, implement, and validate their understanding of the microservice workloads as well as the underlying infrastructure.
RL, on the other hand, is well-suited for learning resource reprovisioning policies, as it provides a tight feedback loop for exploring the action space and generating optimal policies without relying on inaccurate assumptions (i.e., heuristics or rules).
It allows direct learning from actual workload and operating conditions to understand how adjusting low-level resources affects application performance.
In particular, \xxx utilizes the deep deterministic policy gradient (DDPG) algorithm~\cite{lillicrap2015continuous}, which is a \textit{model-free}, \textit{actor-critic} RL framework (shown in \cref{fig:rl-agent}).
Further, \xxx's RL formulation provides two distinct advantages:
\begin{enumerate}
    \item Model-free RL does not need the ergodic distribution of states or the environment dynamics (i.e., transitions between states), which are difficult to model precisely. When microservices are updated, the simulations of state transitions used in model-based RL are no longer valid.
    \item The Actor-critic framework combines policy-based and value-based methods (i.e., consisting of an actor-net and a critic-net as shown in \cref{fig:actor-critic-arch}), and that is suitable for continuous stochastic environments, converges faster, and has lower variance~\cite{grondman2012survey}.
    %Policy-gradient-based methods are more effective in high dimensional or continuous action spaces and tend to converge faster than Q-learning-only methods.
\end{enumerate}

\begin{algorithm}[!t]
\caption{DDPG Training}
%lillicrap2015continuous
\label{algo:training}
\begin{algorithmic}[1]
\State Randomly init $Q_w(s, a)$ and $\pi_\theta(a|s)$ with weights $w$ \& $\theta$.
\State Init target network $Q'$ and $\pi'$ with $w'\leftarrow w$ \& $\theta'\leftarrow \theta$
\State Init replay buffer $\mathcal{D} \gets \varnothing$
\For{episode = $1$, $M$}
    \State Initialize a random process $\mathcal{N}$ for action exploration
    \State Receive initial observation state $s_1$
    \For{$t = 1, T$}
        % according to the current policy and exploration noise
        \State Select and execute action $a_t = \pi_\theta(s_t) + \mathcal{N}_t$
        %\State Execute action $a_t$
        \State Observe reward $r_t$ and new state $s_{t+1}$
        \State Store transition $(s_t, a_t, r_t, s_{t+1})$ in $\mathcal{D}$
        % random mini-batch
        \State Sample $N$ transitions $(s_i, a_i, r_i, s_{i+1})$ from $\mathcal{D}$
        % \State Set $y_i = r_i + \gamma Q'_{w'}(s_{i+1}, \pi'_{\theta'}(a|s_{i+1}))$
        \State Update critic by minimizing the loss $\mathcal{L}(w)$
        \State Update actor by sampled policy gradient $\nabla_\theta J$
        % update the target network
        \State $w' \leftarrow \gamma w + (1-\gamma)w'$
        \State $\theta' \leftarrow \gamma\theta + (1-\gamma)\theta'$
    \EndFor
\EndFor
%\Statex $y_i = r_i + \gamma Q'_{w'}(s_{i+1}, \pi'_{\theta'}(s_{i+1}))$
%\Statex $^1\mathcal{L}(w) = \frac{1}{N}\sum_i(y_i-Q_w(s_i, a_i))^2$
%\Statex $^2\nabla_\theta J = \frac{1}{N}\sum_i \nabla_a Q_w(s=s_i,a=\pi(s_i))\nabla_\theta \pi_\theta(s=s_i)$
\end{algorithmic}
\end{algorithm}

\textbf{Learning the Optimal Policy.}
%\xxx uses an actor-critic technique~\cite{grondman2012survey} to learn the policy $\pi_\theta$.
DDPG's policy learning is an actor-critic approach.
Here the ``critic'' estimates the \textit{value function} (i.e., the expected value of cumulative discounted reward under a given policy), and the ``actor'' updates the policy in the direction suggested by the critic.
The critic's estimation of the expected return allows the actor to update with gradients that have lower variance, thus speeding up the learning process (i.e., achieving convergence).
We further assume that the actor and critic are represented as deep neural networks.
%and train them using the deep deterministic policy gradient method (DDPG)~\cite{lillicrap2015continuous}.
DDPG also solves the issue of dependency between samples and makes use of hardware optimizations by introducing a \textit{replay buffer}, which is a finite-sized cache $\mathcal{D}$ that stores transitions ($s_t$,$a_t$,$r_t$,$s_{t+1}$).
Parameter updates are based on a mini-batch of size $N$ sampled from the reply buffer.
The pseudocode of the training algorithm is shown in Algorithm \ref{algo:training}.
RL training proceeds in episodes and each episode consists of $T$ time steps. At each time step, both actor and critic neural nets are updated once.

%In \xxx the critic uses the popular \textit{Q-Learning} approach~\cite{watkins1992q}.
In the critic, the value function $Q_w(s_t,a_t)$ with parameter $w$ and its corresponding loss function are defined as:
%Here, instead of learning the value function for a state, we learn the value function for a state-action pair $(s, a)$, i.e., $Q_w(s, a) = \mathbb{E}_\pi[G_t|s_t = s, a_t = a]$.
\vspace{-3pt}
\[Q_w(s_t,a_t) = \mathbb{E}[r(s_t, a_t) + \gamma Q_w(s_{t+1}, \pi(s_{t+1}))]\]
%It aims to minimize the mean squared loss 
\vspace{-3pt}
\[\mathcal{L}(w) = \frac{1}{N}\sum_i(r_i + \gamma Q'_{w'}(s_{i+1}, \pi'_{\theta'}(s_{i+1}))-Q_w(s_i, a_i))^2.\]
The target networks $Q'_{w'}(s,a)$ and $\pi'_{\theta'}(s)$ are introduced in DDPG to mitigate the problem of instability and divergence when one is directly implementing deep RL agents.
In the actor component, DDPG maintains a parametrized actor function $\pi_\theta(s)$, which specifies the current policy by deterministically mapping states to a specific action.
The actor is updated as follows: 
\vspace{-3pt}
\[\nabla_\theta J = \frac{1}{N}\sum_i \nabla_a Q_w(s=s_i,a=\pi(s_i))\nabla_\theta \pi_\theta(s=s_i).\]
%$\mathcal{L}(\pi_\theta) = -\sum_{t=0}^{T}Q^{\pi_\theta}(s_t,a_t)log(\pi_\theta(a_t|s_t))$.

\textbf{Problem Formulation.}
To estimate resources for a microservice instance, we formulate a sequential decision-making problem which can be solved by the above RL framework.
Each microservice instance is deployed in a separate container with a tuple of resource limits $RLT = (RLT_{cpu}, RLT_{mem}, RLT_{llc}, RLT_{io}, RLT_{net})$, since we are considering CPU utilization, memory bandwidth, LLC capacity, disk I/O bandwidth, and network bandwidth as our resource model.\footnote{The resource limit for the CPU utilization of a container is the smaller of $\hat{R_i}$ and the number of threads $\times$ 100.}
% we are not considering for memory capacity because it has little effect on performance
% instead, when memory capaciy is not enough the OOM error will lead to container restart
This limit for each type of resource is predetermined (usually overprovisioned) before the microservices are deployed in the cluster and later controlled by \xxx.

At each time step $t$, utilization $RU_t$ for each type of resource is retrieved using performance counters as telemetry data in \circled{1}.
In addition, \xxx's Extractor also collects current latency, request arrival rate, and request type composition (i.e., percentages of each type of request).
Based on these measurements, the RL agent calculates the states listed in \cref{tab:network_in_out} and described below.
\begin{itemize}
    \item \textit{SLO maintenance ratio} ($SM_t$) is defined as \texttt{SLO\_latency/ current\_latency} if the microservice instance is determined to be the culprit. If no message arrives, it is assumed that there is no SLO violation ($SM_t = 1$).
    \item \textit{Workload changes} ($WC_t$) is defined as the ratio of the arrival rates at the current and previous time steps.
    \item \textit{Request composition} ($RC_t$) is defined as a unique value encoded from an array of request percentages by using \texttt{numpy.ravel\_multi\_index()}~\cite{numpy}.
\end{itemize}

For each type of resources $i$, there is a predefined resource upper limit $\hat{R_i}$ and a lower limit $\uhat{R}_i$ (e.g., the CPU time limit cannot be set to 0).
The actions available to the RL-agent is to set $RLT_i\in [\hat{R_i}, \uhat{R}_i]$.
If the amount of resource reaches the total available amount, then a scale-out operation is needed.
Similarly, if the resource limit is below the lower bound, a scale-in operation is needed.
The CPU resources serve as one exception to the above procedure: it would not improve the performance if the CPU utilization limit were higher than the number of threads created for the service.

The goal of the RL agent is, given a time duration $t$, to determine an optimal policy $\pi_t$ that results in as few SLO violations as possible (i.e., $\min_{\pi_t} SM_t$) while keeping the resource utilization/limit as high as possible (i.e., $\max_{\pi_t} RU_t/RLT_t$).
Based on both objectives, the reward function is then defined as $r_t = \alpha\cdot SM_t\cdot|\mathcal{R}| + (1-\alpha)\cdot\sum_i^{|\mathcal{R}|} RU_i/RLT_i$, where $\mathcal{R}$ is the set of resources.

% if the bottleneck is CPU but currently the number of threads is equal to the number of cores,
%  - either notify the microservice owner that they could use multi-threading
%  - at the same time scale-out to rectify the SLO violation

%However, in our case directly imposing exploration noise to the output action performs poorly because resource constraints cause invalid exploration.
%To deal with this issue, we apply parameter space noise~\cite{plappert2017parameter}.
%Instead of adding noise to the action space, we perturb the policy using additive Gaussian noise for exploration and train the non-perturbed network on this data by replaying it.
%Experimental results show that introducing parameter space noise solves the resource constraints issue and makes training converge fast.

\textbf{Transfer Learning.}
Using a tailored RL agent for every microservice instead of using the shared RL agent should improve resource reprovisioning efficiency, as the model would be more sensitive to application characteristics and features.
However, such an approach is hard to justify in practice (i.e., for deployment) because of the time required to train such tailored models for user workloads, which might have significant churn.
\xxx addresses the problem of rapid model training by using transfer learning in the domain of RL~\cite{taylor2009transfer,taylor2008autonomous,celiberto2010using}, whereby agents for SLO violation mitigation can be trained for either the general case (i.e., any microservices) or the specialized case (i.e., ``transferred'' to the behavior of individualized microservices).
The pre-trained model used in the specialized case is called the base model or the source model.
That approach is possible because prior understanding of a problem structure helps one solve similar problems quickly, with the remaining task being to understand the behavior of updated microservice instances.
Related work on base model selection and task similarity can be found in ~\cite{taylor2009transfer,taylor2008autonomous}, but the base model that \xxx uses for transfer learning is always the RL model learned in the general case because it has been shown in evaluation to be comparable with specialized models.
We demonstrate the efficacy of transfer learning in our evaluation described in \cref{sec:eval}.
The RL model that \xxx uses is designed to scale since both the state space and the action space are independent of the size of the application or the cluster.
In addition to having the general case RL agent, the \xxx framework also allows for the deployment of specialized per-microservice RL agents.

\begin{table}[!t]
    \centering
    \caption{State-action space of the RL agent.}
    \label{tab:network_in_out}
    \begin{threeparttable}
    \resizebox{\columnwidth}{!}{%
        \begin{tabular}{p{8cm}}
            \toprule
            \rowcolor{gray!15}
            \multicolumn{1}{c}{\textbf{State ($s_t$)}}\\
            SLO Maintenance Ratio ($SM_t$), Workload Changes ($WC_t$),
            Request Composition ($RC_t$),
            Resource Utilization ($RU_t$)\\
            \rowcolor{gray!15}
            \multicolumn{1}{c}{\textbf{Action Space ($a_t$)}}\\
            Resource Limits $RLT_i(t),i \in \{\text{CPU, Mem, LLC, IO, Net}\}$\\
            %RL_m(t), RL_{lc}(t), RL_i(t), RL_n(t)
            \bottomrule
        \end{tabular}%
        }
    %\begin{tablenotes}
    %    \footnotesize
    %    \item[*] SLO Viol. Ratio = SLO Latency / Current Latency \texttt{if} Violations \texttt{else} 1
    %\end{tablenotes}
    \end{threeparttable}
\end{table}

\begin{table}[!t]
    \centering
    \caption{RL training parameters.}
    \resizebox{\columnwidth}{!}{%
    \begin{tabular}{ll}
    \toprule
    \textbf{Parameter} & \textbf{Value} \\ \midrule
    \# Time Steps $\times$ \# Minibatch & 300 $\times$ 64 \\
    Size of Replay Buffer & $10^5$ \\
    Learning Rate & Actor ($3\times 10^{-4}$), Critic ($3\times 10^{-3}$) \\
    Discount Factor & 0.9 \\
    Soft Update Coefficient & $2\times 10^{-3}$ \\
    Random Noise & $\mu$ (0), $\sigma$ (0.2) \\ 
    Exploration Factor & $\epsilon$ (1.0), $\epsilon$-decay ($10^{-6}$) \\ \bottomrule
    \end{tabular}%
    }
    \label{table:rl-params}
\end{table}

\textbf{Implementation Details.}
We implemented the DDPG training algorithm and the actor-critic networks using \texttt{PyTorch}~\cite{pytorch}.
The critic net contains two fully connected hidden layers with 40 hidden units, all using ReLU activation function.
The first two hidden layers of the actor net are fully connected and both use ReLU as the activation function while the last layer uses Tanh as the activation function.
The actor network has 8 inputs and 5 outputs, while the critic network has 23 inputs and 1 output.
The actor and critic networks are shown in \cref{fig:actor-critic-arch}, and their inputs and outputs are listed in \cref{tab:network_in_out}.
We chose that setup because adding more layers and hidden units does not increase performance in our experiments with selected microservice benchmarks; instead, it slows down training speed significantly.
Hyperparameters of the RL model are listed in \cref{table:rl-params}.
%We choose the number of time steps in each episode to be 300 but for initial stages, we terminate the RL exploration early so that the agent can reset and try again from the initial state.
We set the time step for training the model to be 1 second, which is sufficient for action execution (see Table \ref{table:op-time}).
%We use a replay buffer of size $10^5$ and set the size of the mini-batch to be 64.
%Actor learning rate, critic learning rate, and the discount factor are set to be $3\times 10^{-4}$, $3\times 10^{-3}$, and 0.9 respectively.
The latencies of each RL training update and inference step are 73 $\pm$ 10.9 ms and 1.2 $\pm$ 0.4 ms, respectively.
The average CPU and memory usage of the Kubernetes pod during the training stage are 210 millicores and 192 Mi, respectively.

\subsection{Action Execution}
\label{sec:action-execution}

\xxx's Deployment Module, i.e., \circled{5}, verifies the actions generated by the RL agent and executes them accordingly.
Each action on scaling a specific type of resource is limited by the total amount of the resource available on that physical machine.
\xxx assumes that machine resources are unlimited and thus does not have admission control or throttling.
If an action leads to oversubscribing of a resource, then it is replaced by a scale-out operation.
\begin{itemize}
    \item \textit{CPU Actions:} Actions on scaling CPU utilization are executed through modification of \texttt{cpu.cfs\_period\_us} and \texttt{cpu.cfs\_quota\_us} in the \texttt{cgroups} CPU subsystem.
    \item \textit{Memory Actions:} We use Intel MBA~\cite{intelmba} and Intel CAT~\cite{intelcat} technologies to control the memory bandwidth and LLC capacity of containers, respectively.\footnote{Our evaluation on IBM Power systems (see \cref{sec:eval}) did not use these actions because of a lack of hardware support.
    OS support or software partitioning mechanisms~\cite{lin2008gaining, arch-support-for-cmp} can be applied; we leave that to future work.}
    \item \textit{I/O Actions:}  For I/O bandwidth, we use the \texttt{blkio} subsystem in \texttt{cgroups} to control input/output access to disks.
    \item \textit{Network Actions:} For network bandwidth, we use the Hierarchical Token Bucket (HTB)~\cite{htb} queueing discipline in Linux Traffic Control.
    Egress \texttt{qdisc}s can be directly shaped by using HTB.
    Ingress \texttt{qdisc}s are redirected to the virtual device \texttt{ifb} interface and then shaped through the application of egress rules.
\end{itemize}

\iffalse
\begin{figure}[ht]
    \centering
    \includegraphics[width=\linewidth]{figs/span-graph-darker.pdf}
    \caption{Simplified example of the \textit{Span} (left) and execution history graph for an end-to-end microservice request. Individual \textit{Spans} and service instances are categorized into two groups: parallel and sequential. One exception is called background \textit{Span}, which is out of the user request, e.g., \texttt{write-to-timeline} service in \texttt{compose-post} request.}
    \label{fig:span}
\end{figure}
\fi

\iffalse
\begin{equation}
\begin{aligned}
    &
    \begin{aligned}
        Q(s_t,a_t) \leftarrow & Q(s_t,a_t) +\\
        &\alpha(r_{t+1}+\gamma \max_a Q(s_{t+1},a)-Q(s_t,a_t))
    \end{aligned}\\
    &
    \begin{aligned}
        \theta_t \leftarrow & \theta_t +\\
        &\alpha(Q(s_{t},a_t)\cdot \nabla_\theta \ln{\pi(a_t|s_t,\theta_t)})
    \end{aligned}
\end{aligned}
\end{equation}
\fi

%Our implementation of a prototype of \xxx (\cref{implementation:orchestrator}) is based on Kubernetes~\cite{container:k8s} which adapts a centralized master-slave architecture.
%It consists of a tracing coordinator and extractor for tracing coordination, data movement and critical instance extraction, an RL-agent for resource estimation, and a deployment module for action verification and execution.
%We implemented the inter-component communication based on protocol buffers~\cite{protobuf}.
%We also describe the implementation of the performance anomaly injector (\cref{sec:fault-injector}) for \xxx design, training and assessment.

\begin{table}[!t]
    \centering
    \caption{Types of performance anomalies injected to microservices causing SLO violations.}
    \resizebox{\columnwidth}{!}{%
    \begin{tabular}{ll}
    \toprule
    \textbf{Performance Anomaly Types} & \textbf{Tools/Benchmarks} \\ \midrule
    Workload Variation & \texttt{wrk2}~\cite{wrk2} \\
    Network Delay & \texttt{tc}~\cite{tc} \\
    CPU Utilization & iBench~\cite{delimitrou2013ibench}, \texttt{stress-ng}~\cite{stressng} \\
    LLC Bandwidth \& Capacity & iBench, \texttt{pmbw}~\cite{pmbw} \\
    Memory Bandwidth & iBench~\cite{delimitrou2013ibench}, \texttt{pmbw}~\cite{pmbw} \\
    I\slash O Bandwidth & Sysbench~\cite{sysbench} \\
    Network Bandwidth & \texttt{tc}~\cite{tc}, Trickle~\cite{trickle} \\ \bottomrule
    \end{tabular}%
    }
    \label{table:fault-injector}
\end{table}

\subsection{Performance Anomaly Injector}
\label{sec:fault-injector}

We accelerate the training of the machine learning models in \xxx's Extractor and the RL agent through performance anomaly injections.
The injection provides the ground truth data for the SVM model, as the injection targets are controlled and known from the campaign files.
It also allows the RL agent to quickly span the space of adverse resource contention behavior (i.e., the exploration-exploitation trade-off in RL).
That is important, as real-world workloads might not experience all adverse situations within a short training time.
We implemented a performance anomaly injector, i.e., \circled{6}, in which the injection targets, type of anomaly, injection time, duration, patterns, and intensity are configurable.
The injector is designed to be bundled into the microservice containers as a file-system layer; the binaries incorporated into the container can then be triggered remotely during the training process.
The injection campaigns (i.e., how the injector is configured and used) for the injector will be discussed in \cref{sec:eval}.
The injector comprises seven types of performance anomalies that can cause SLO violations. They are listed in \cref{table:fault-injector} and described below.

\textbf{Workload Variation.}
We use an HTTP benchmarking tool \texttt{wrk2} as the workload generator. It performs multithreaded, multiconnection HTTP request generation to simulate client-microservice interaction. The request arrival rate and distribution can be adjusted to break the predefined SLOs.

\textbf{Network Delay.}
We use Linux traffic control (\texttt{tc}) to add simulated delay to network packets. Given the mean and standard deviation of the network delay latency, each network packet is delayed following a normal distribution.

\textbf{CPU Utilization.}
We implement the CPU stressor based on iBench and \texttt{stree-ng} to exhaust a specified level of CPU utilization on a set of cores by exercising floating point, integer, bit manipulation and control flows.

\textbf{LLC Bandwidth \& Capacity.}
We use iBench and \texttt{pmbw} to inject interference on the Last Level Cache (LLC).
% The benchmark mines the \texttt{/proc/cpuinfo} of the operating system.
For bandwidth, the injector performs streaming accesses in which the size of the accessed data is tuned to the parameters of the LLC.
For capacity, it adjusts intensity based on the size and associativity of the LLC to issue random accesses that cover the LLC capacity.

\textbf{Memory Bandwidth.}
We use iBench and \texttt{pmbw} to generate memory bandwidth contention.
It performs serial memory accesses (of configurable intensity) to a small fraction of the address space.
Accesses occur in a relatively small fraction of memory in order to decouple the effects of contention in memory bandwidth from contention in memory capacity.
%Containers are sometimes forced to terminate when out of memory capacity, so we are not interested in memory capacity problems.

\textbf{I/O Bandwidth.}
We use Sysbench to implement the file I\slash O workload generator.
It first creates test files that are larger than the size of system RAM.
Then it adjusts the number of threads, read/write ratio, and sleeping/working ratio to meet a specified level of I/O bandwidth.
We also use Tricle for limiting the upload/download rate of a specific microservice instance.

\textbf{Network Bandwidth.}
We use Linux traffic control (\texttt{tc}) to limit egress network bandwidth.
For ingress network bandwidth, an intermediate function block (\texttt{ifb}) pseudo interface is set up, and inbound traffic is directed through that.  In that way, the inbound traffic then becomes schedulable by the egress \texttt{qdisc} on the \texttt{ifb} interface, so the same rules for egress can be applied directly to ingress.

    \section{Evaluation}
\label{sec:eval}

\subsection{Experimental Setup}
\label{eval:setup}

\textbf{Benchmark Applications.}
We evaluated \xxx on a set of end-to-end interactive and responsive real-world microservice benchmarks: 
\begin{enumerate*}[label=(\roman*)]
\item DeathStarBench~\cite{gan2019open}, consisting of \textit{Social Network}, \textit{Media Service}, and \textit{Hotel Reservation} microservice applications, and 
\item Train-Ticket~\cite{zhou2018fault}, consisting of the \textit{Train-Ticket Booking Service}
\end{enumerate*}.
\textit{Social Network} implements a broadcast-style social network with unidirectional follow relationships whereby users can publish, read, and react to social media posts.
\textit{Media Service} provides functionalities such as reviewing, rating, renting, and streaming movies.
\textit{Hotel Reservation} is an online hotel reservation site for browsing hotel information and making reservations.
\textit{Train-Ticket Booking Service} provides typical train-ticket booking functionalities, such as ticket inquiry, reservation, payment, change, and user notification.
These benchmarks contain 36, 38, 15, and 41 unique microservices, respectively; cover all workflow patterns (see \cref{sec:critical-path}); and use various programming languages including Java, Python, Node.js, Go, C/C++, Scala, PHP, and Ruby.
All microservices are deployed in separate Docker containers.

\textbf{System Setup.}
We validated our design by implementing a prototype of \xxx that used Kubernetes~\cite{container:k8s} as the underlying container orchestration framework.
We deployed the four microservice benchmarks with \xxx separately on a Kubernetes cluster of 15 two-socket physical nodes without specifying any anti-colocation rules.
Each server consists of 56--192 CPU cores and RAM that varies from 500 GB to 1000 GB.
Nine of the servers use Intel {x86} Xeon E5s and E7s processors, while the remaining ones use IBM {ppc64} Power8 and Power9 processors.
All machines run Ubuntu 18.04.3 LTS with Linux kernel version 4.15.
% All experiments are performed on both Intel and IBM machines. 
% However, due to lack of hardware support for cache-level partitioning in IBM CPUs, we only demonstrate resource partitioning only on the Intel machines.

\textbf{Load Generation.}
We drove the services with various open-loop asynchronous workload generators~\cite{wrk2} to represent an active production environment~\cite{sriraman2019softsku,ueda2016workload,chen2017workload}.
We uniformly generated workloads for every request type across all microservice benchmarks.
The parameters for the workload generators were the same as those for DeathStarBench (which we applied to Train-Ticket as well), and varied from predictable constant, diurnal, distributions such as Poisson, to unpredictable loads with spikes in user demand.
The workload generators and the microservice benchmark applications were never co-located (i.e., they executed on different nodes in the cluster). To control the variability in our experiments, we disabled all other user workloads on the cluster.  
% These workload generators (microservice users) are running on different machines from the Kubernetes cluster where microservices are running on.
% During these experiments, there are only microservice benchmark workloads. All other workloads are disabled through Docker swarm.

\textbf{Injection and Comparison Baselines.}
We used our performance anomaly injector (see \cref{sec:fault-injector}) to inject various types of performance anomalies into containers uniformly at random with configurable injection timing and intensity.
Following the common way to study resource interference, our experiments on SLO violation mitigation with anomalies were designed to be comprehensive by covering the worst-case scenarios, given the random and nondeterministic nature of shared-resource interference in production environments~\cite{patros2016investigating,dean2013tail}.
Unless otherwise specified, (i) the anomaly injection time interval was in an exponential distribution with $\lambda=0.33 s^{-1}$, and (ii) the anomaly type and intensity were selected uniformly at random.
We implemented two baseline approaches: (a) the Kubernetes autoscaling mechanism~\cite{k8sautoscaling} and (b) an AIMD-based method~\cite{studli2015modified,gevros2004distributed} to manage resources for each container.
Both approaches are rule-based autoscaling techniques.

\begin{figure*}[!t]
    \begin{subfigure}[b]{0.25\textwidth}  
        \centering 
        \includegraphics[width=\textwidth]{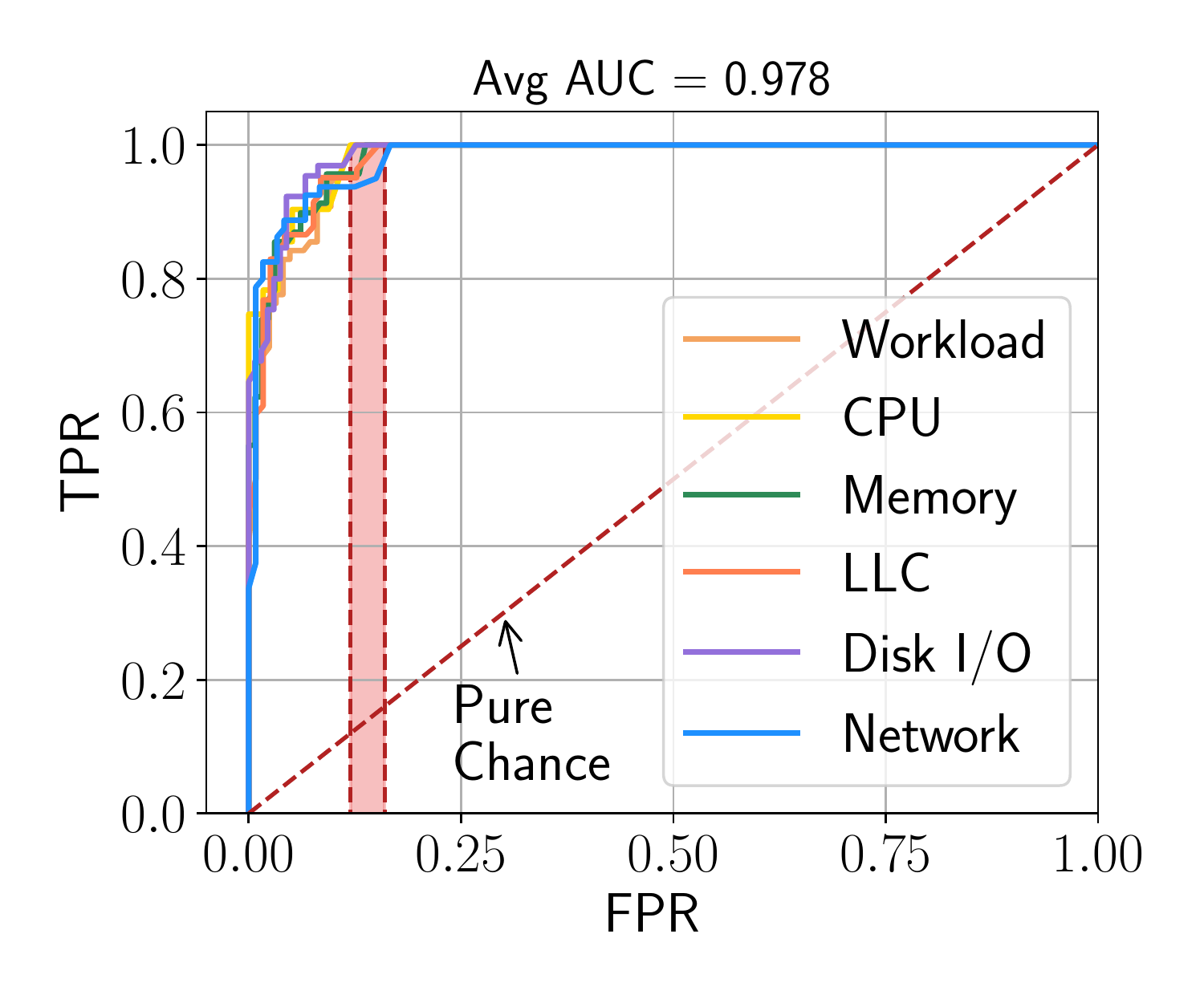}
        \label{fig:roc-single}
        \vspace{-20pt}
        \caption{ROC under single-anomaly.}
    \end{subfigure}
    %\quad
    \begin{subfigure}[b]{0.35\textwidth}  
        \centering 
        \includegraphics[width=\textwidth]{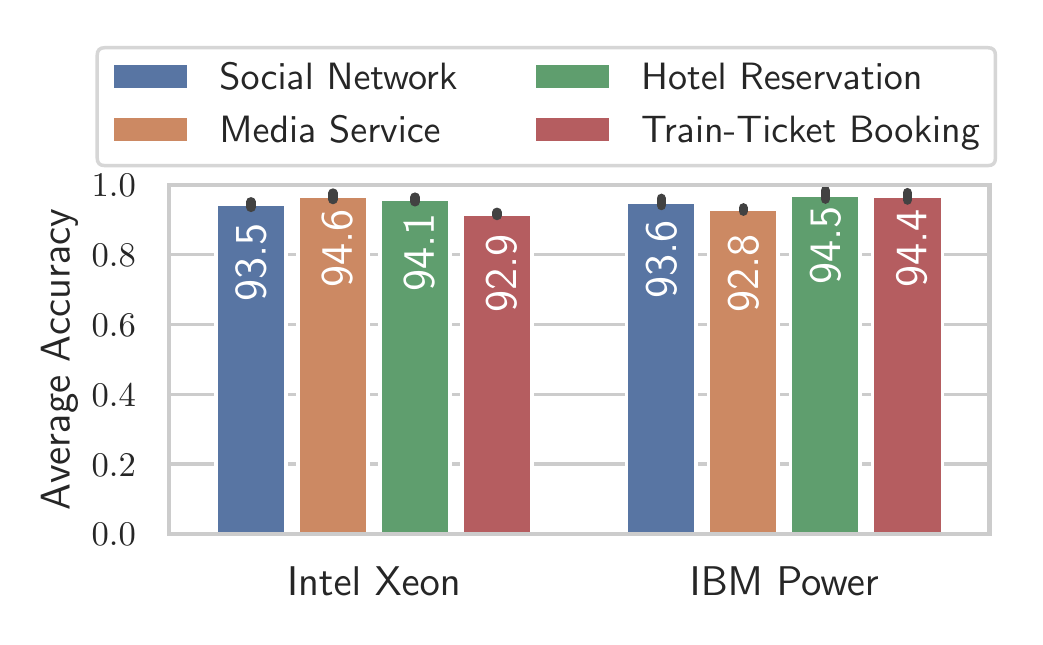}
        \label{fig:roc-avg}
        \vspace{-20pt}
        \caption{Average accuracy under multi-anomaly.}
    \end{subfigure}
    \begin{subfigure}[b]{0.4\textwidth}
        \centering 
        \includegraphics[width=\textwidth]{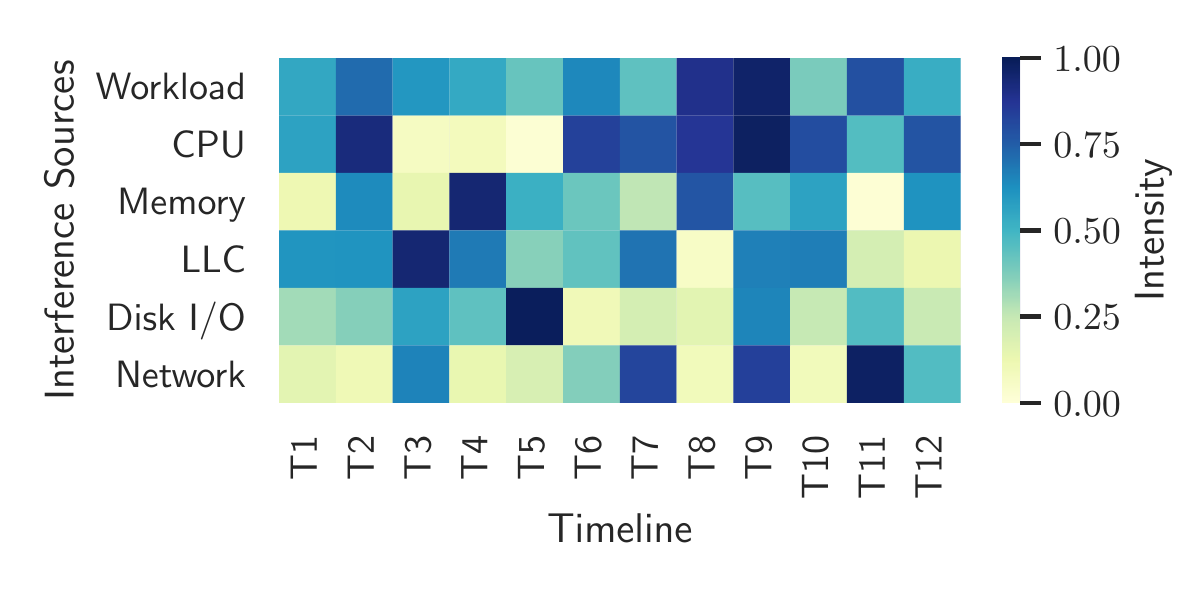}
        \label{fig:roc-sources}
        \vspace{-20pt}
        \caption{Anomaly injection intensity and timing.} 
    \end{subfigure}
    \caption{Critical Component Localization Performance: (a) ROC curves for detection accuracy; (b) Variation of localization accuracies across processor architectures; (c) Anomaly-injection intensity, types, and timing.}
    \label{fig:roc}
\end{figure*}

\subsection{Critical Component Localization}
\label{eval:rca}
Here, we use the techniques presented in \cref{sec:critical-path,sec:critical-component} to study the effectiveness of \xxx in identifying the microservices that are most likely to cause SLO violations. 
%under both single and multiple performance anomaly injections.
% roc curve for single-fault injection

\textbf{Single anomaly localization.}
We first evaluated how well \xxx localizes the microservice instances that are responsible for SLO violations under different types of single-anomaly injections.
For each type of performance anomaly and each type of request, we gradually increased the intensity of injected resource interference and recorded end-to-end latencies. The intensity parameter was chosen uniformly at random between [start-point, end-point], where the start-point is the intensity that starts to trigger SLO violations, and the end-point is the intensity when either the anomaly injector has consumed all possible resources or over 80\% of user requests have been dropped or returned time.
% Then the intensity parameter of this experiment is chosen from [start-point, end-point] uniformly at random.
\cref{fig:roc}(a) shows the receiver operating characteristic (ROC) curve of root cause localization.
The ROC curve captures the relationship between the false-positive rate (x-axis) and the true-positive rate (y-axis).
The closer to the upper-left corner the curve is, the better the performance.
We observe that the localization accuracy of \xxx, when subject to different types of anomalies, does not vary significantly.
In particular, \xxx's Extractor module achieved near $100\%$ true-positive rate, when the false-positive rate was between $[0.12, 0.16]$.

% learning curve: total reward vs. iterations
% avg recover time vs. iterations (compare against k8s autoscaling and AIMD)
\begin{figure}[!t]
    \begin{subfigure}[b]{0.5\columnwidth}
        \centering
        \includegraphics[width=\columnwidth]{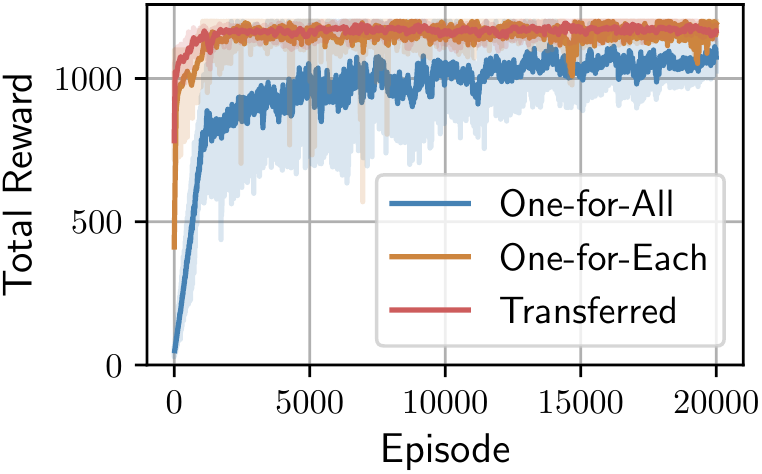}
        \caption{Total reward.}
        \label{fig:rl-training-curve}
    \end{subfigure}
    \hfill
    \begin{subfigure}[b]{0.49\columnwidth}
        \centering
        \includegraphics[width=\columnwidth]{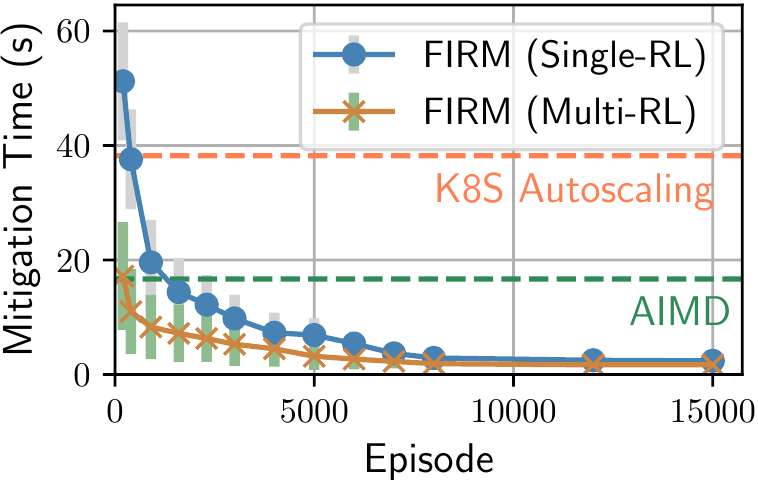}
        \caption{SLO mitigation time.}
        \label{fig:rl-recovery-time}
    \end{subfigure}
    \caption{Learning curve showing total reward during training and SLO mitigation performance.}
    \label{fig:rl-learning-curve}
    %\caption{SLO Violation Mitigation Time Comparison Againt Kubernetes Autoscaling, and AIMD-based Method with Increasing Number of Training Episodes.}
\end{figure}

% avg accuracy for multi-fault injection across apps across platforms
\textbf{Multi-anomaly localization.} 
There is no guarantee that only one resource contention will happen at a time under dynamic datacenter workloads~\cite{sriraman2020accelerometer,sriraman2018mu,hazelwood2018applied,gmach2007workload} and
therefore we also studied the container localization performance under multi-anomaly injections and compared machines with two different processor ISAs ({x86} and {ppc64}).
An example of the intensity distributions of all the anomaly types used in this experiment are shown in \cref{fig:roc}(c).
The experiment was divided into time windows of 10 s, i.e., $T_i$ from \cref{fig:roc}(c)).
At each time window, we picked the injection intensity of each anomaly type uniformly at random with range [0,1].
Our observations are reported in \cref{fig:roc}(b).
The average accuracy for localizing critical components in each application ranged from 92\% to 94\%.
The overall average localization accuracy was $93\%$ across four microservice benchmarks.
Overall, we observe that the accuracy of the Extractor did not differ between the two sets of processors.

% end to end
\begin{figure*}[!t]
    \centering
    \includegraphics[width=\linewidth]{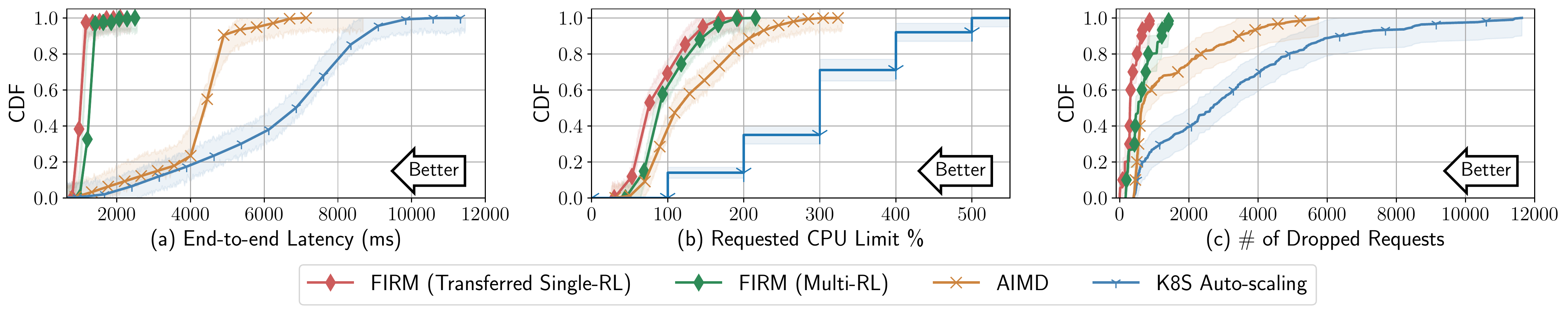}
    \caption{Performance comparisons (CDFs) of end-to-end latency, requested CPU limit, and the number of dropped requests.}
    \label{fig:latency-eval}
    % more accurately
    %\xxx achieves $6.9\times$ and $11.5\times$ lower 99th-percentile end-to-end latency on average than baseline approaches, leading to $9.8\times$ and $16.7\times$ less SLO violations for orchestrating microservices under resource contention scenarios. Meanwhile, \xxx lowers requested CPU limit by 35.9-67\%.
\end{figure*}

\subsection{RL Training \& SLO Violation Mitigation}
\label{eval:training}

To understand the convergence behavior of \xxx's RL agent, we trained three RL models that were subjected to the same sequence of performance anomaly injections (described in \cref{eval:setup}).
The three RL models are: (i) a common RL agent for all microservices (one-for-all), (ii) a tailored RL agent for a particular microservice (one-for-each), and (iii) a transfer-learning-based RL agent.
RL training proceeds in episodes (iterations).
We set the number of time steps in a training episode to be 300 (see \cref{table:rl-params}), but for the initial stages, we terminate the RL exploration early so that the agent could reset and try again from the initial state. % cumulative learning
We did so because the initial policies of the RL agent are unable to mitigate SLO violations.
Continuously injecting performance anomalies causes user requests to drop, and thus only a few request traces were generated to feed the agent.
As the training progressed, the agent improved its resource estimation policy and could mitigate SLO violations in less time.
At that point (around 1000 episodes), we linearly increased the number of time steps to let the RL agent interact with the environment longer before terminating the RL exploration and entering the next iteration.
%This makes the resource management problem more challenging and steers the agent to learn stronger policies over time.

%To study the convergence behavior of the RL agent 
We trained the abovementioned three RL models on the Train-Ticket benchmark. We studied the generalization of the RL model by evaluating the end-to-end performance of \xxx on the DeathStarBench benchmarks. Thus, we used DeathStarBench as a validation set in our experiments.
%we show three learning curves: a common RL agent for all microservices (one-for-all), a tailored RL agent (one-for-each), and a transfer-learning-based RL agent 
%sharing the parameter of the learned common RL agent
\cref{fig:rl-learning-curve}(a) shows that as the training proceeded, the agent was getting better at mitigation, and thus the moving average of episode rewards was increasing.
The initial steep increase benefits from early termination of episodes and parameter exploration.
Transfer-learning-based RL converged even faster (around 2000 iterations\footnote{1000 iterations correspond to roughly 30 minutes with each iteration consisting of 300 time steps.}) because of parameter sharing.
The one-for-all RL required more iterations to converge (around 15000 iterations) and had a slightly lower total reward (6\% lower compared with one-for-each RL) during training.

In addition, higher rewards, for which the learning algorithm explicitly optimizes, correlate with improvements in SLO violation mitigation (see \cref{fig:rl-learning-curve}(b)).
For models trained in every 200 episodes, we saved the checkpoint of parameters in the RL model.
Using the parameter, we evaluated the model snapshot by injecting performance anomalies (described in \cref{eval:setup}) continuously for one minute and observed when SLO violations were mitigated.
\cref{fig:rl-learning-curve}(b) shows that \xxx with either a single-RL agent (one-for-all) or a multi-RL agent (one-for-each) improved with each episode in terms of the SLO violation mitigation time. The starting policy at iteration 0--900 was no better than the Kubernetes autoscaling approach, but after around 2500 iterations, both agents were better than either Kubernetes autoscaling or the AIMD-based method.
Upon convergence, \xxx with a single-RL agent achieved a mitigation time of 1.7 s on average, which outperformed the AIMD-based method by up to 9$\times$ and Kubernetes autoscaling by up to 30$\times$ in terms of the time to mitigate SLO violations.
% 1.7s 16.7s 37.2s

\subsection{End-to-End Performance}
\label{eval:recovery}
% \vspace{-4pt}

Here, we show the end-to-end performance of \xxx and its generalization by further evaluating it on DeathStarBench benchmarks based on the hyperparameter tuned during training with the Train-Ticket benchmark.
To understand the 10--30$\times$ improvement demonstrated above, we measured the 99th percentile end-to-end latency when the microservices were being managed by the two baseline approaches and by \xxx.
\cref{fig:latency-eval}(a) shows the cumulative distribution of the end-to-end latency.
We observed that the AIMD-based method, albeit simple, outperforms the Kubernetes autoscaling approach by $1.7\times$ on average and by $1.6\times$ in the worst case.
In contrast, \xxx: 
\begin{enumerate}
    \item Outperformed both baselines by up to $6\times$ and $11\times$, which leads to $9\times$ and $16\times$ fewer SLO violations; 
    \item Lowered the overall requested CPU limit by 29--62\%, as shown in \cref{fig:latency-eval}(b), and increased the average cluster-level CPU utilization by up to 33\%; and
    \item Reduced the number of dropped or timed out user requests by up to 8$\times$ as shown in \cref{fig:latency-eval}(c).
\end{enumerate}
\xxx can provide these benefits because it detects SLO violations accurately and addresses resource contention before SLO violations can propagate.
% through end-to-end RL training
By interacting with dynamic microservice environments under complicated loads and resource allocation scenarios,
\xxx's RL agent dynamically learns the policy, and hence outperforms heuristics-based approaches.

%\begin{figure*}
%    \centering
%    
%    \begin{minipage}{0.59\textwidth}
%        \centering
%        \captionsetup{type=table} %% tell latex to change to table
%        \vspace{20pt}
%        \resizebox{\linewidth}{!}{%
%        \begin{tabular}{lrrrrrrr}
%            \toprule
%            \multirow{2}{*}{\textbf{Operation}} & \multicolumn{5}{c}{\textbf{Partition (Scale Up/Down)}} & \multicolumn{2}{c}{\textbf{Pod/Container}} \\
%            \cmidrule(r){2-6}\cmidrule(l){7-8}
%             & \textbf{CPU} & \textbf{Memory} & \textbf{LLC} & \textbf{I/O} & \textbf{Network} & \textbf{Warm-Start} & \textbf{Cold-Start} \\ 
%            \midrule
%            Mean ($ms$) & 2.1 & 42.4 & 39.8 & 2.3 & 12.3 & 45.7 & 2050.8 \\
%            SD ($ms$) & 0.3 & 11.0 & 9.2 & 0.4 & 1.1 & 6.9 & 291.4 \\
%            \bottomrule
%            \end{tabular}%
%        }
%        \vspace{3pt}
%        \caption{Average latency for resource management operations.}
%        \label{table:op-time}
%    \end{minipage}
%\end{figure*}

%\begin{minipage}{0.4\textwidth}
%        \centering
%        \includegraphics[width=1.7in]{figs/k8s-load-balancing.pdf}
%        \vspace{-5pt}
%        \caption{Kubernetes load-balancing latency.}
%        \label{fig:k8s-load-balancing}
%    \end{minipage}
%    \hfill
    \section{Discussion}
\label{sec:discussion}

%\textbf{Critical Path Changes.}
%Through experiments with performance anomaly injections, we observed that the major reason for critical path changes on serving a request is variance in individual service latency.
%We attribute this to three causes:
%(i) Microservice components are deployed on different physical machines with different amounts of resource contention. % there could be large variance in individual latency.
%(ii) Different services are sensitive to different types of resource interference. % --> variance under different resource contention
%(iii) Some service is on different execution paths (e.g., \texttt{composePost} in \cref{fig:microservice-overview}) serving different endpoints, but each endpoint has different sensitivity.
%The lesson we learned here is to decouple each individual microservice component as much as possible, so that the issue of different sensitivity across endpoints could be mitigated.

% talk about perfect spike prediction is impossible
% talk about if operation delay time > spike duration, then there's no way to mitigate this

\textbf{Necessity and Challenges of Modeling Low-level Resources.}
Recall from \cref{sec:background} that modeling of resources at a fine granularity is necessary, as it allows for better performance without overprovisioning.
It is difficult to model the dependence between low-level resource requirements and quantifiable performance gain while dealing with uncertain and noisy measurements~\cite{ott2018hardware,wang2016hardware}. 
\xxx addresses the issue by modeling that dependency in an RL-based feedback loop,
which automatically explores the action space to generate optimal policies without human intervention.
% lower level noizy, not clear how changes in this level lead to app performance (give more changes in cahce -> performance? maybe or maybe not) hard to understand parameters

\textbf{Why a Multilevel ML Framework?}
A model of the states of all microservices that is fed as the input to a single large ML model~\cite{yang2019miras,prachitmutita2018auto} leads to 
\begin{enumerate*}[label=(\roman*)] 
\item state-action space explosion issues that grow with the number of microservices, thus increasing the training time; and
\item dependence between the microservice architecture and the ML-model, which sacrifices the generality.
\end{enumerate*}
\xxx addresses those problems by incorporating a two-level ML framework.
The first level ML model uses SVM to filter the microservice instances responsible for SLO violations, thereby reducing the number of microservices that need to be considered in mitigating SLO violations.
%The second ML model learns resource management policies on per-microservice basis instead of considering the whole application state using reinforcement learning.  
That enables the second level ML model, the RL agent, to be trained faster and removes dependence on the application architecture.
That, in turn, helps avoid RL model reconstruction/retraining.
%by learning resource management policies on a per-microservice basis instead of considering the whole application state.
% More importantly, could drive the research of underlying performance models by replacing SVM model and relying on RL's automatic feedback loop..

\begin{table}[!t]
    \centering
    \caption{Avg. latency for resource management operations.}
    \label{table:op-time}
    \resizebox{\linewidth}{!}{%
    \begin{tabular}{lrrrrrrr}
        \toprule
        \multirow{2}{*}{\textbf{Operation}} & \multicolumn{5}{c}{\textbf{Partition (Scale Up/Down)}} & \multicolumn{2}{c}{\textbf{Container Start}} \\
        \cmidrule(r){2-6}\cmidrule(l){7-8}
         & \textbf{CPU} & \textbf{Mem} & \textbf{LLC} & \textbf{I/O} & \textbf{Net} & \textbf{Warm} & \textbf{Cold} \\ 
        \midrule
        Mean ($ms$) & 2.1 & 42.4 & 39.8 & 2.3 & 12.3 & 45.7 & 2050.8 \\
        Std Dev ($ms$) & 0.3 & 11.0 & 9.2 & 0.4 & 1.1 & 6.9 & 291.4 \\
        \bottomrule
        \end{tabular}%
    }
\end{table}

\textbf{Lower Bounds on Manageable SLO Violation Duration for \xxx.}
As shown in Table \ref{table:op-time}, the operations to scale resources for microservice instances take 2.1--45.7 ms.
Thus, that is the minimum duration of latency spikes that any RM approach can handle.
For transient SLO violations, which last shorter than the minimum duration, the action generated by \xxx will always miss the mitigation deadline and can potentially harm overall system performance.
Worse, it may lead to oscillations between scaling operations.
%is useless because the transient violation has gone before the action takes effect.
Predicting the spikes before they happen, and proactively taking mitigation actions can be a solution.
However, it is a generally-acknowledged difficult problem, as microservices are dynamically evolving, in terms of both load and architectural design, which is subject to our future work.

\textbf{Limitations.}
\xxx has several limitations that we plan to address in future work. First, \xxx currently focuses on resource interference caused by real workload demands. However, \xxx lacks the ability to detect application bugs or misconfigurations, which may lead to failures such as memory leak. Allocating more resources to such microservice instances may harm the overall resource efficiency. Other sources of SLO violations, including global resource sharing (e.g., network switches or global file systems) and hardware causes (e.g., power-saving energy management), are also beyond \xxx's scope.
Second, the scalability of \xxx is limited by the maximum scalability of the centralized graph database, and the boundary caused by the network traffic telemetry overhead. (Recall the lower bound on the SLO violation duration.)
Third, we plan to implement \xxx's tracing module based on side-car proxies (i.e., service meshes)~\cite{chandramouli2020building} that minimizes application instrumentation and has wider support of programming languages.

    \section{Related Work}
\label{sec:related}

SLO violations in cloud applications and microservices are a popular and well-researched topic.
We categorize prior work into two buckets: root cause analyzers and autoscalers.
Both rely heavily on the collection of tracing and telemetry data.

\textbf{Tracing and Probing for Microservices.}
Tracing for large-scale microservices (essentially distributed systems)
%is an old and challenging problem.
%There are extensive number of work in this area.
%Some work~\cite{fonseca2007x,sigelman2010dapper,tracing-zipkin,tracing-jaeger,tracing-lightstep,tracing-instana,tracing-skywalking,chow2014mystery} requires application-level instrumentation. X-Trace~\cite{fonseca2007x} is one of the earliest work which does cross-layer, cross-application tracing to reconstruct application's task tree.
%Sharing conceptual similarities with X-Trace, Google's distributed tracing infrastructure Dapper~\cite{sigelman2010dapper} makes use of sampling and restricts instrumentation to reduce development burden and runtime overhead.
helps understand the path of a request as it propagates through the components of a distributed system.
Tracing requires either application-level instrumentation~\cite{fonseca2007x,sigelman2010dapper,tracing-zipkin,tracing-jaeger,tracing-lightstep,tracing-instana,tracing-skywalking,chow2014mystery,kaldor2017canopy}
%(e.g. Dapper, X-Trace, Jaeger, Zipkin, etc.),
or middleware/OS-level instrumentation~\cite{liu2019jcallgraph,barham2003magpie,chen2002pinpoint,thalheim2017sieve} (e.g., Sieve~\cite{thalheim2017sieve} utilizes a kernel module \textit{sysdig}~\cite{sysdig} which provides system calls as an event stream containing tracing information about the monitored process to a user application).

\textbf{Root Cause Analysis.}
% Large-scale and complex distributed systems such as microservices are susceptible to anomalies that break predefined SLOs~\cite{wu2020microrca,thalheim2017sieve,liu2019jcallgraph}.
%Given the large scale, heterogeneous services, and complex dependencies, root causes are extremely hard to localize and diagnose.
%On the other hand, 
A large body of work~\cite{shah2018root,jayathilaka2017performance,lin2018microscope,weng2018root,liu2019jcallgraph,thalheim2017sieve,wu2020microrca,chen2002pinpoint,gan2019seer,kaleidoscope} provides promising evidence that data-driven diagnostics help detect performance anomalies and analyze root causes.
For example, Sieve~\cite{thalheim2017sieve} leverages Granger causality to correlate performance anomaly data series with particular metrics as potential root causes.
Pinpoint~\cite{chen2002pinpoint} runs clustering analysis on Jaccard similarity coefficient to determine the components that are mostly correlated with the failure.
Microscope~\cite{lin2018microscope} and MicroRCA~\cite{wu2020microrca} are both designed to identify abnormal services by constructing service causal graphs that model anomaly propagation and by inferring causes using graph traversal or ranking algorithms~\cite{jeh2003scaling}.
%Microscope~\cite{lin2018microscope} is designed to identify abnormal services with a ranked list of possible root causes in microservices environments.
%It does this by constructing a service causal graph, and then traversing the graph backward from the violation node to infer the causes of performance problems.
%MicroRCA~\cite{wu2020microrca} infers root causes in real time by correlating application performance symptoms with corresponding system resource utilization.
%It generates an attributed graph that model anomaly propagation across services and machines, and then ranks each node to locate root causes using PageRank~\cite{jeh2003scaling}.
Seer~\cite{gan2019seer} uses deep learning to learn spatial and temporal patterns that translate to SLO violations.
%and the massive amount of tracing data
However, none of these approaches addresses the dynamic nature of microservice environments (i.e., frequent microservice updates and deployment changes), which
% addition, replication, and updates
require costly model reconstruction or retraining.
% these algorithms on graph operations does not scale with the microservice complexity
% seer does not work for replica set
%More importantly, the above mentioned approaches do not address the dynamicity in microservice environments.
%Frequent microservice changes require burdensome model reconstruction or neural network retraining.

\textbf{Autoscaling Cloud Applications.}
Current techniques for autoscaling cloud applications can be categorized into four groups~\cite{qu2018auto,lorido2014review}:
(a) rule-based (commonly offered by cloud providers~\cite{aws:autoscaling,azure:autoscaling,gcloud:autoscaling}),
(b) time series analysis (regression on resource utilization, performance, and workloads),
(c) model-based (e.g., queueing networks), or
(d) RL-based.
Some approaches combine several techniques.
For instance, Auto-pilot~\cite{rzadca2020autopilot} combines time series analysis and RL algorithms to scale the number of containers and associated CPU/RAM.
Unfortunately, when applied to microservices with large scale and complex dependencies, independent scaling of each microservice instance results in suboptimal solutions (because of critical path intersection and insight 2 in \cref{sec:background}), and it is difficult to define sub-SLOs for individual instances.
%cannot directly be applied to microservices, which itself contains hundreds or thousands of heterogeneous components.
Approaches for autoscaling microservices or distributed dataflows~\cite{yu2019microscaler,gias2019atom,prachitmutita2018auto,kalavri2018three,yang2019miras} make scaling decisions on the number of replicas and/or container size without considering low-level shared-resource interference.
%either result in sub-optimal resource provisioning solution, or does not model the nature of microservices that each service can independently scale.
% Most importantly they do not consider low-level shared-resource interference when planning.
% It only considers the number of replicas and CPU shares.
%ATOM~\cite{gias2019atom} constructs the layered queueing network model of the application,
%then uses computational optimization to solve a non-linear mixed-integer program.
%By doing so, ATOM dynamically control the number of replicas for each microservices component and its associated container CPU share.
% slow - in total 2 min
% there are cases no need to scale but only to partition some shared resources
%Microscaler~\cite{yu2019microscaler} proposes a criterion named service power to determine the service(s) that are needed to scale.
%It decides the service size by combining a Bayesian Optimization model and a step-by-step heuristic model.
ATOM~\cite{gias2019atom} and Microscaler~\cite{yu2019microscaler} do so by using a combination of queueing network- and heuristic-based approximations.
%both control the number of replicas and/or associated container CPU share by 
ASFM~\cite{prachitmutita2018auto} uses recurrent neural network activity to predict workloads and translates application performance to resources by using linear regression.
%scales CPU cores and memory capacity using resource scaling optimization algorithm based on predicted workload with neural networks
%, recurrent neural network and .
% Workload Predictor (MLP+LSTM) for each service -> Resource Planner (Linear Regression on # of CPU & Memory) -> Feedback
Streaming and data-processing scalers like DS2~\cite{kalavri2018three} and MIRAS~\cite{yang2019miras} leverage explicit application-level modeling and apply RL to represent the resource-performance mapping of operators and their dependencies.
%knowledge of the data-flow graph, the computational dependencies among operators, and estimates each operator’s true processing and output rates.

\textbf{Cluster Management.}
The emergence of cloud computing motivates the prevalence of cloud management platforms that provide services such as monitoring, security, fault tolerance, and performance predictability.
Examples include Borg~\cite{verma2015large}, Mesos~\cite{hindman2011mesos}, Tarcil~\cite{delimitrou2015tarcil}, Paragon~\cite{delimitrou2013paragon}, Quasar~\cite{delimitrou2014quasar}, Morpheus~\cite{jyothi2016morpheus}, DeepDive~\cite{novakovic2013deepdive}, and Q-clouds~\cite{nathuji2010q}.
In this paper, we do not address the problem of cluster orchestration.
\xxx can work in conjunction with those cluster management tools to reduce SLO violations.

    \section{Conclusion}
\label{sec:conclusion}
We propose \textit{\xxx}, an ML-based, fine-grained resource management framework that addresses SLO violations and resource underutilization in microservices.
% \xxx builds on three key ideas: identifying critical paths in execution history graphs using online performance analysis, localizing problematic microservices using performance variability on per-critical-path and per-instance levels, and mitigating SLO violations using RL-based dynamic reprovisioning.
\xxx uses a two-level ML model, one for identifying microservices responsible for SLO violations, and the other for mitigation. The combined ML model reduces SLO violations up to 16$\times$ while reducing the overall CPU limit by up to 62\%. 
%\xxx consists of an online critical component analysis which is adaptive to dynamically changing microservice environments, and an RL agent taking care of bottleneck analysis and resource estimation.
% We implemented \xxx in our local cluster and validated it against four real-world microservice applications.
% Through end-to-end RL training, \xxx reduces SLO violations up to 16.7\texttimes{} while reducing overall CPU limit by up to 62.3\%.
Overall, \xxx enables fast mitigation of SLOs by using efficient resource provisioning, which benefits both cloud service providers and microservice owners.
\xxx is open-sourced at \url{https://gitlab.engr.illinois.edu/DEPEND/firm.git}.

% This work could drive the research on underlying models relating fine-grained low-level resources to application performance - future work
    \section{Acknowledgment}

We thank the OSDI reviewers and our shepherd, Rebecca Isaacs, for their valuable comments that improved the paper. We appreciate K. Atchley, F. Rigberg, and J. Applequist for their insightful comments on the early drafts of this manuscript.
This research was supported in part by the U.S. Department of Energy, Office of Science, Office of Advanced Scientific Computing Research, under award No. 2015-02674.
This work is partially supported by the National Science Foundation (NSF) under grant No. 2029049; by a Sandia National Laboratories\footnote{Sandia National Laboratories is a multimission laboratory managed and operated by National Technology and Engineering Solutions of Sandia, LLC, a wholly owned subsidiary of Honeywell International Inc., for the U.S. Department of Energy's National Nuclear Security Administration under contract DE-NA0003525.} under contract No. 1951381; by the IBM-ILLINOIS Center for Cognitive Computing Systems Research (C3SR), a research collaboration that is part of the IBM AI Horizon Network; and by Intel and NVIDIA through equipment donations.
Any opinions, findings, and conclusions or recommendations expressed in this material are those of the authors and do not necessarily reflect the views of the NSF, Sandia National Laboratories, IBM, NVIDIA, or, Intel.
Saurabh Jha is supported by a 2020 IBM PhD fellowship.
    {
        \bibliographystyle{plain}
        \bibliography{main}
    }
    %\input{01000-appendix}
%%%%%%%%%%%%%%%%%%%%%%%%%%%%%%%%%%%%%%%%%%%%%%%%%%%%%%%%%%%%%%%%%%%%%%%%%%

%%%%%%%%%%%%%%%%%%%%%%%%%%%%%%%%%%%%%%%%%%%%%%%%%%%%%%%%%%%%%%%%%%%%%%%%%%
\end{document}